\documentclass[journal]{IEEEtran}
\usepackage{cite}
\usepackage{amsfonts,amssymb,stmaryrd}
\usepackage{amsbsy}
\usepackage{graphicx}
\usepackage{latexsym}
\usepackage{amsmath,mathrsfs}
\usepackage{psfrag}
\usepackage{subfigure}
\newtheorem{Def}{Definition}
\newtheorem{Eg}{Example}
\newtheorem{Prop}{Proposition}
\newtheorem{Thm}{Theorem}
\newtheorem{Lem}{Lemma}
\newtheorem{Coro}{Corollary}
\newtheorem{Rem}{Remark}

\newcommand{\A}{\mathcal{A}}
\newcommand{\B}{\mathcal{B}}
\newcommand{\E}{\mathcal{E}}
\newcommand{\F}{\mathcal{F}}
\newcommand{\K}{\mathcal{K}}
\newcommand{\M}{\mathcal{M}}
\newcommand{\LL}{\mathcal{L}}
\newcommand{\OO}{\mathcal{O}}
\newcommand{\PP}{\mathcal{P}}

\begin{document}
\title{Observability and Decentralized Control of Fuzzy Discrete Event Systems}
\author{Yongzhi Cao and Mingsheng Ying%
\thanks{This work was supported by the National Foundation of
Natural Sciences of China under Grants 60496321, 60321002, and
60273003, by the Chinese National Key Foundation Research
$\binampersand$ Development Plan (2004CB318108), and by the Key
Grant Project of Chinese Ministry of Education under Grant 10403.}%
\thanks{The authors are with the State Key Laboratory of Intelligent Technology and
Systems, Department of Computer Science and Technology, Tsinghua
University, Beijing 100084, China (e-mail:
caoyz@mail.tsinghua.edu.cn, yingmsh@mail.tsinghua.edu.cn).}}

\markboth{CAO AND YING: Observability and Decentralized Control of Fuzzy Discrete Event
Systems}{CAO AND YING: Observability and Decentralized Control of Fuzzy Discrete Event Systems}
\maketitle

\begin{abstract}Fuzzy discrete event systems as a generalization of (crisp) discrete event systems
have been introduced in order that it is possible to effectively
represent uncertainty, imprecision, and vagueness arising from the
dynamic of systems. A fuzzy discrete event system has been modelled
by a fuzzy automaton; its behavior is described in terms of the
fuzzy language generated by the automaton. In this paper, we are
concerned with the supervisory control problem for fuzzy discrete
event systems with partial observation. Observability, normality,
and co-observability of crisp languages are extended to fuzzy
languages. It is shown that the observability, together with
controllability, of the desired fuzzy language is a necessary and
sufficient condition for the existence of a partially observable
fuzzy supervisor. When a decentralized solution is desired, it is
proved that there exist local fuzzy supervisors if and only if the
fuzzy language to be synthesized is controllable and co-observable.
Moreover, the infimal controllable and observable fuzzy
superlanguage, and the supremal controllable and normal fuzzy
sublanguage are also discussed. Simple examples are provided to
illustrate the theoretical development.
\end{abstract}


\section{Introduction}
\PARstart{D}{iscrete} event systems (DES) are systems whose state
space is discrete and whose state can only change as a result of
asynchronously occurring instantaneous events over time. Such
systems have been successfully applied to provide a formal
treatment of many man-made systems such as communication systems,
networked systems, manufacturing systems, and automated traffic
systems. The behavior of a DES is described in terms of the
sequences of events involved. Supervisory control of DES pioneered
by Ramadge and Wonham \cite{RW87} and subsequently extended by
many other researchers (see, for example, \cite{RW89},
\cite{CL99}, and the bibliographies therein) provides a framework
for designing supervisors for controlling the behavior of DES.

Usually, a DES is described by finite state automaton with events
as input alphabets, and the behavior is thus the language accepted
by the automaton. It is worth noting that such a model can only
process crisp state transitions. In other words, no uncertainty
arises in the state transitions of the model. There are, however,
many situations such as mobile robots in an unstructured
environment \cite{MG02}, intelligent vehicle control \cite{R03},
and wastewater treatment \cite{WSEAD00}, in which the state
transitions of some systems are always somewhat imprecise,
uncertain, and vague. A convincing example given in \cite{LY01}
and \cite{LY02} is a patient's condition, where the change of the
condition from a state, say ``excellent", to another, say ``fair",
is obviously imprecise, since it is hard to measure exactly the
change.

Vagueness, imprecision, and uncertainty are typical features of
most of the complex systems. It is well known that the methodology
of fuzzy sets first proposed by Zadeh \cite{Z65} is a good tool
for coping with imprecision, uncertainty, and vagueness. To
capture significant uncertainty appearing in states and state
transitions of DES, Lin and Ying have incorporated fuzzy set
theory together with DES and thus have extended crisp DES to fuzzy
DES by applying fuzzy finite automaton model \cite{LY01,LY02}.
Under the framework of fuzzy DES, Lin and Ying have discussed
state-based observability and some optimal control problems.
Excitingly, the first application of fuzzy DES has recently been
reported by Ying $et$ $al.$ in \cite{Y04}, where fuzzy DES are
used to handle treatment planning for HIV/AIDS patients. It is
worth noting that fuzzy finite automata and fuzzy languages have
some important applications to many other fields such as clinical
monitoring and pattern recognition (see, for example,
\cite{MM02}).

As a continuation of the works \cite{LY01} and \cite{LY02}, we
have developed supervisory control theory for fuzzy DES modelled
by (maxmin) fuzzy automata in \cite{CY}. The behavior of such
systems is described by their generated fuzzy languages.
Informally, a fuzzy language consists of certain event strings
associated with membership grade. The membership grade of a string
can be interpreted as the possibility degree to which the system
in its initial state and with the occurrence of events in the
string  may enter another state. Although strings in a
probabilistic language \cite{P71},\cite{GKM99,KG01} are also
endowed with weight, it should be pointed out that fuzzy languages
are different from probabilistic languages in semantics: the
weight in a fuzzy context describes the membership grade (namely,
uncertainty) of a string, while the weight of a string in a
probabilistic context reflects a frequency of occurrence. This
difference appeals for distinct control action and can also
satisfy diverse applications.

For control purposes, the set of events is partitioned into two
disjoint subsets of controllable and uncontrollable events, as
usually done in crisp DES. Control is exercised by a fuzzy
supervisor that disables controllable events with certain degrees
in the controlled system, also called a plant, so that the
closed-loop system of supervisor and plant exhibits a
pre-specified desired fuzzy language. The controllability of fuzzy
languages has been introduced in  \cite{CY} as a necessary and
sufficient condition for the existence of a fuzzy supervisor that
achieves a desired specification for a given fuzzy DES. A similar
controllability condition based on different underlying
constituents of fuzzy automata and uncontrollable event set has
also been given independently by Qiu in \cite{Q05}.

In \cite{CY} and  \cite{Q05}, both Qiu and we have restricted our
attention to the problem of centralized control under full
observation. There are, however, some systems in which supervisors
cannot ``see" or ``observe" all the events. For example, in a
diagnosis on a disease, if we regard the evolvement of the disease
as a system with some physicians as supervisors, it is a fact that
the physicians cannot see all the events that result in the
evolvement of the disease. Unobservability of events arises
principally from the limitations of the sensors attached to the
systems and the distributed nature of some systems such as
manufacturing systems and communication networks where events at
some locations are not seen at other locations. For crisp DES, the
issue of partial observation has been extensively studied in both
centralized and decentralized control (see, for example,
\cite{CDFV88,LW88,LW88D,LW90,RW92}). However, in terms of fuzzy
DES, this issue is investigated only as a global property of
systems in \cite{LY01,LY02}; a further study is desired.

The purpose of this paper is to develop these earlier works
\cite{LY01,LY02,CY,Q05} and address the supervisory control problem
for fuzzy DES with partial observation. This requires that we have
to deal with the presence of unobservable events in addition to the
presence of uncontrollable events. We are mainly concerned with what
controlled behavior can be achieved when controlling a fuzzy DES
with a partially observable fuzzy supervisor and decentralized fuzzy
supervisors, respectively.

The contribution of this paper is as follows.

 1). To characterize the class of fuzzy languages achievable under the partially observable architecture,
in Section III we introduce the notions of observability and strong
observability of fuzzy languages which are generalized versions of
observability of crisp languages. We show that there exists a
partially observable fuzzy supervisor synthesizing a desired fuzzy
language if and only if the fuzzy language is controllable and
observable. The property that observable fuzzy languages are closed
under arbitrary intersections leads to the existence of infimal
controllable and observable fuzzy superlanguage.

2). Decentralized supervisory control with global specification
has been investigated in Section IV. By generalizing
co-observability of crisp languages to fuzzy languages, a
necessary and sufficient condition for the existence of local
fuzzy supervisors that achieve a given legal fuzzy language is
derived.

3). In order to obtain an approximation for a given fuzzy language
by using controllable and observable sublanguages, normality of
fuzzy languages which is stronger than observability is defined in
Appendix I. Some properties of normal fuzzy languages and the
relation between normality, observability, and controllability are
also presented.

Besides the sections mentioned above, Section II provides the
necessary preliminaries; Section V addresses an illustrative
example; Section VI concludes the work presented and identifies
some future research directions. The paper also contains two
appendices: Appendix I is devoted to the normality of fuzzy
languages and Appendix II shows the proofs of theorems,
propositions, and lemmas.

\section{Preliminaries}
In the first subsection, we will briefly recall a few basic facts
on the supervisory control of crisp DES under partial observation.
For a detailed introduction to the supervisory control theory,
readers may refer to \cite{CL99}. The second subsection is devoted
to the background on the supervisory control of fuzzy DES.
\subsection{Observability and Decentralized Control of Crisp DES}
Let $E$ denote the finite set of events, and $E^*$ denote the set
of all finite sequences of events, or stings, in $E$, including
the empty string $\epsilon$. A string $\mu\in E^*$ is a prefix of
a string $\omega\in E^*$ if there exists $\nu\in E^*$ such that
$\mu\nu=\omega$. In this case, we write $\mu\leq\omega$. The
length of a string $\omega$ is denoted by $|\omega|$. Any subset
of $E^*$ is called a language over $E$. The prefix closure of a
language $L$, denoted by $\overline{L}$,  consists of the set of
strings which are prefixes of strings in $L$. A language $L$ is
said to be prefix closed if $L=\overline{L}$.

A crisp DES, or plant, is usually described by a deterministic automaton: $G=(Q,E,\delta,q_0)$,
where $Q$ is a set of states with the initial state $q_0$, $E$ is a set of events, and
$\delta:Q\times E\rightarrow Q$ is a (partial) transition function. The function $\delta$ is
extended to $\delta:Q\times E^*\rightarrow Q$ in the obvious way. In a logical model of a DES, we
are interested in the strings of events that the system can generate. Thus the behavior of a DES is
modelled as a prefix closed language $L(G)=\{s\in E^*:\delta(q_0,s)\mbox{ is defined}\}$ over the
event set $E$.

Recall that associated with the system is a set $E_c$ of events that
can be disabled, and there is a set of events, $E_o$, that can be
observed by partial observation supervisors. The sets of
uncontrollable and unobservable events are denoted by
$E_{uc}=E\backslash E_{c}$ and $E_{uo}=E\backslash E_{o}$,
respectively. To represent the fact that a partial observation
supervisor has only a partial observation of strings in $L(G)$, a
natural projection operator $P:E^*\rightarrow E^*_o$ is used. Recall
that $P(\epsilon)=\epsilon$ and $P(sa)=P(s)P(a)$ for any $s\in E^*$
and $a\in E$, where $P(a)=a$ if $a\in E_o$, and otherwise
$P(a)=\epsilon$. A (partial observation) supervisor is a map
$S_P:P[L(G)]\rightarrow2^E$ such that $S_P[P(s)]\supseteq E_{uc}$
for any $s\in L(G)$.

The language generated by the controlled system, denoted by $L(S_P/G)$, is defined inductively as
follows:

1) $\epsilon\in L(S_P/G);$

2) $[s\in L(S_P/G)$, $sa\in L(G)$, and $a\in S_P[P(s)]]\Leftrightarrow[sa\in L(S_P/G)].$

Let us recall two key notions in crisp DES. \vspace{0.1cm}
\begin{Def}A prefix-closed language $K\subseteq L(G)$ is said to be
{\it controllable} \cite{RW87} (with respect to $L(G)$ and $E_{uc}$) if $KE_{uc}\cap L(G)\subseteq
K$; it is called {\it observable} \cite{LW88} (with respect to $L(G)$, $P$, and $E_{c}$) if for any
$s,s'\in K$ and any $a\in E_c$, $$[P(s)=P(s'),\ sa\in K,\mbox{ and }s'a\in L(G)]\Rightarrow[s'a\in
K].$$\label{Dobser}
\end{Def}

Given a nonempty and prefix-closed language $K\subseteq L(G)$, it
has been shown in \cite{CDFV88} and \cite{LW88} that there exists
a partial observation supervisor $S_P$ such that $L(S_P/G)=K$ if
and only if $K$ is controllable and observable.

If the plant $G$ is physically distributed, then it is desirable to have decentralized supervisors,
where each supervisor is able to control a certain set of events and is able to observe certain
other events. For the sake of simplicity, let us recall the decentralized supervisory control with
only two local supervisors.

Let $E_{ic},E_{io}\subseteq E,i=1,2$, be the local controllable and observable event sets,
respectively. Let $P_i:E^*\rightarrow E^*_{io}$ be the corresponding natural projection. The local
supervisor $S_{iP}$ (or simply $S_i$), $i=1,2$, is given by the map $S_{i}:P_i[L(G)]\rightarrow2^E$
that satisfies $S_{i}[P_i(s)]\supseteq E\backslash E_{ic}$ for any $s\in L(G)$.

The language $L(S_{1}\wedge S_{2}/G)$ generated by the system $G$ under the joint supervision of
$S_1$ and $S_2$ is defined inductively by
\begin{enumerate}
    \item $\epsilon\in L(S_{1}\wedge S_{2}/G)$;
    \item $[s\in L(S_{1}\wedge S_{2}/G),\ sa\in L(G),
    \mbox{ and }a\in S_1[P_1(s)]\cap S_{2}[P_2(s)]]\Rightarrow[sa\in L(S_{1}\wedge S_{2}/G)].$
\end{enumerate}

To state the existence condition for local supervisors, it is
necessary to introduce co-observability, which is defined below.
\vspace{0.2cm}
\begin{Def}A prefix-closed language $K\subseteq L(G)$ is {\it co-observable} \cite{RW92} (with respect to $L(G)$,
$P_i$, and $E_{ic}$, $i=1,2$), if for all $s,s',s''\in K$ subject to $P_1(s)=P_1(s')$ and
$P_2(s)=P_2(s'')$, the following hold:
\begin{enumerate}
    \item if $a\in E_{1c}\cap E_{2c}$, $sa\in L(G)$, and $s'a,s''a\in K$, then
$sa\in K$;
    \item if $a\in E_{1c}\backslash E_{2c}$, $sa\in L(G)$, and $s'a\in K$, then
$sa\in K$;
    \item if $a\in E_{2c}\backslash E_{1c}$, $sa\in L(G)$, and $s''a\in K$, then
$sa\in K$.
\end{enumerate}
\end{Def}
\vspace{0.2cm}

 Given a nonempty and prefix-closed language
$K\subseteq L(G)$, it has been shown by Rudie and Wonham in
\cite{RW92} that there exist local supervisors $S_1$ and $S_2$
such that $L(S_{1}\wedge S_{2}/G)=K$ if and only if $K$ is
controllable and co-observable.

\subsection{Fuzzy DES and Its Controllability}
In this subsection, we recall the model of fuzzy DES and its
supervisory control under full observation.

For later need, let us first review some notions and notation on fuzzy set theory. Each {\it fuzzy
subset} (or simply {\it fuzzy set}), $\A$, is defined in terms of a relevant universal set $X$ by a
function assigning to each element $x$ of $X$ a value $\A(x)$ in the closed unit interval $[0,1]$.
Such a function is called a {\it membership function}, which is a generalization of the
characteristic function associated to a crisp set; the value $\A(x)$ characterizes the degree of
membership of $x$ in $\A$.

The {\it support} of a fuzzy set $\A$ is a crisp set defined as
$\mbox{supp}(\A)=\{x:\A(x)>0\}$. Whenever $\mbox{supp}(\A)$ is a
finite set, say $\mbox{supp}(\A)=\{x_1,x_2,\cdots,x_n\}$, then
fuzzy set $\A$ can be written in Zadeh's notation as
follows:$$\A=\frac{\A(x_1)}{x_1}+\frac{\A(x_2)}{x_2}+\cdots+\frac{\A(x_n)}{x_n}.$$

We denote by $\F(X)$ the set of all fuzzy subsets of $X$. For any $\A,\B\in \F(X)$, we say that
$\A$ is contained in $\B$ (or $\B$ contains $\A$), denoted by $\A\subseteq\B$, if $\A(x)\leq\B(x)$
for all $x\in X$. We say that $\A=\B$ if and only if $\A\subseteq\B$ and $\B\subseteq\A$. A fuzzy
set is said to be {\it empty} if its membership function is identically zero on $X$. We use $\OO$
to denote the empty fuzzy set.

For any family $\lambda_i,\ i\in I$, of elements of $[0,1]$, we write $\vee_{i\in I}\lambda_i$ or
$\vee\{\lambda_i:i\in I\}$ for the supremum of $\{\lambda_i:i\in I\}$, and $\wedge_{i\in
I}\lambda_i$ or $\wedge\{\lambda_i:i\in I\}$ for its infimum. In particular, if $I$ is finite, then
$\vee_{i\in I}\lambda_i$ and $\wedge_{i\in I}\lambda_i$ are the greatest element and the least
element of $\{\lambda_i:i\in I\}$, respectively.

Now, we are able to introduce the model of fuzzy DES. A fuzzy DES
is modelled by a fuzzy automaton which is known as maxmin
automaton in some mathematical literature \cite{S68}, \cite{KL79}
and is somewhat different from max-product automata used in
\cite{LY01,LY02}.

\vspace{0.2cm}
\begin{Def}A {\it fuzzy automaton} is a four-tuple $G=(Q, E,\delta, q_0)$, where:
\begin{itemize}
    \item $Q$ is a crisp (finite or infinite) set of states;
    \item $E$ is a finite set of events;
    \item $q_0\in Q$ is the initial state;
    \item $\delta$ is a function from $Q\times E\times Q$ to $[0,1]$, called a fuzzy transition function.
\end{itemize}
\end{Def}

\vspace{0.2cm} For any $p,\ q\in Q$ and $a\in E$, we can interpret
$\delta(p,a,q)$ as the possibility degree to which the automaton
in state $p$ and with the occurrence of event $a$ may enter state
$q$. Contrast to crisp DES, an event in fuzzy DES may take the
system to more than one states with different degrees. The concept
of fuzzy automata is a natural generalization of nondeterministic
automata. The major difference between fuzzy automata and
nondeterministic automata is: in a nondeterministic automaton,
$\delta(p,a,q)$ is either $1$ or $0$, so the possibility degrees
of existing transitions are the same, but if we work with a fuzzy
automaton, they may be different.

An {\it extended fuzzy transition function} from $Q\times E^*\times Q$ to $[0,1]$, denoted by the
same notation $\delta$, can be defined inductively as follows:
\begin{displaymath}
\delta(p,\epsilon,q)=\left\{ \begin{array}{ll}
1, & \textrm{if $q=p$}\\
0, & \textrm{otherwise}\qquad\qquad
\end{array} \right.
\end{displaymath}
$$\delta(p,\omega a,q)=\vee_{r\in
Q}(\delta(p,\omega,r)\wedge\delta(r,a,q))
$$
for all $\omega\in E^{*}$ and $a\in E$. The possibility degrees of
transitions here are given by the operation $max$ $min$; this is
the main difference between maxmin automata and max-product
automata used in \cite{LY01} and \cite{LY02}.

Further, the language $\LL(G)$ generated by $G$, called {\it fuzzy language}, is defined as a fuzzy
subset of $E^*$ and given by $$\LL(G)(\omega)=\vee_{q\in Q}\delta(q_0,\omega,q).$$ This means that
a string $\omega$ of $E^{*}$ is not necessarily either ``in the fuzzy language $\LL(G)$" or ``not
in the fuzzy language $\LL(G)$"; rather $\omega$ has a membership grade $\LL(G)(\omega)$, which
measures its degree of membership in $\LL(G)$.

If the state set of a fuzzy automaton $G$ is empty, then it yields that $\LL(G)=\OO$, i.e.,
$\LL(G)(\omega)=0$ for all $\omega\in E^*$; otherwise, from the definition we see that $\LL(G)$ has
the following properties:

P1) $\LL(G)(\epsilon)=1;$

P2) $\LL(G)(\mu)\geq \LL(G)(\mu\nu)$ for any $\mu,\nu\in E^*.$

Conversely, given a fuzzy language $\LL$ (over the event set $E$)
satisfying the above properties, we can construct a fuzzy
automaton $G=(Q_\LL,E_\LL,\delta_\LL,q_\LL)$, where
$Q_\LL=\mbox{supp}(\LL)$, $E_\LL=E$, $q_\LL=\epsilon$, and
\begin{displaymath}
\delta_\LL(\mu,a,\nu)=\left\{\begin{array}{ll}
\LL(\nu), & \textrm{if $\nu=\mu a$}\qquad\qquad\qquad\qquad\\
0, & \textrm{otherwise}
\end{array} \right.
\end{displaymath} for all $\mu,\nu\in \mbox{supp}(\LL)$ and $a\in E$.  There is no difficulty to verify
 that $\LL(G)=\LL$. Thus a fuzzy language and a fuzzy DES amount to the same thing in our context, which
 is  analogous to the equivalence of regular expressions and finite
 automata due to Kleene \cite{K56}.

Note that in the rest of the paper, by a fuzzy language we mean the empty fuzzy language $\OO$ or a
fuzzy language that satisfies the above properties P1) and P2), unless otherwise specified. In
particular, the supports of such fuzzy languages are prefix closed by the property P2). We use
$\F\LL$ to denote the set of all fuzzy languages over $E$. More explicitly,
$\F\LL=\{\A\in\F(E^*):\A=\OO\textrm{ or }\A \textrm{ satisfies P1) and P2)}\}$.

Union and intersection of fuzzy languages can be defined as follows:
$$(\cup_{i\in I}\LL_i)(\omega)=\vee_{i\in I}\LL_i(\omega), \mbox{ for all $\omega\in E^*$,
and}$$
$$(\cap_{i\in I}\LL_i)(\omega)=\wedge_{i\in I}\LL_i(\omega), \mbox{ for all $\omega\in E^*$}.\qquad$$

It has been shown in \cite{CY} that fuzzy languages are closed under arbitrary unions and
intersections, respectively.

The concatenation $\LL_1\LL_2$ of two fuzzy languages $\LL_1$ and $\LL_2$, which is again a fuzzy
language \cite{CY}, is defined by
$$(\LL_1\LL_2)(\omega)=\vee\{\LL_1(\mu)\wedge\LL_2(\nu):\mu,\nu\in E^*\mbox{ and
}\mu\nu=\omega\},$$ for all $\omega\in E^*$.

To model the control action of fuzzy DES, we partition the event set $E$ into {\it controllable}
and {\it uncontrollable} events: $E=E_c\dot\cup E_{uc}$, as usually done in crisp DES. However,
unlike controllable events in crisp DES, a controllable event in fuzzy DES can be disabled with any
degree.

Control is achieved by means of a fuzzy supervisor, which is allowed
to disable any fuzzy sets of controllable events after having
observed an arbitrary string $s\in \mbox{supp}(\LL(G))$. Formally, a
(fully observable) {\it fuzzy supervisor} for $G$ is a map
$S:\mbox{supp}(\LL(G))\rightarrow\F(E)$ such that $S(s)(a)=1$ for
any $s\in \mbox{supp}(\LL(G))$ and $a\in E_{uc}$. The controlled
system is denoted by $S/G$; the behavior of $S/G$ is the fuzzy
language $\LL^S$ obtained inductively as follows:

1) $\LL^S(\epsilon)=1;$

2) $\LL^S(sa)=\LL(G)(sa)\wedge S(s)(a)\wedge\LL^S(s)$ for any $s\in E^*$ and $a\in E$.

To state the controllability of fuzzy languages, let us first introduce a fuzzy subset $\E_{uc}$ of
$E^*$, which is given by
\begin{displaymath}
\E_{uc}(\omega)=\left\{ \begin{array}{ll}
1, & \textrm{if $\omega\in E_{uc}$}\\
0, & \textrm{if $\omega\in E^*\backslash E_{uc}$.}
\end{array} \right.
\end{displaymath}

\begin{Def}A fuzzy language $\K\subseteq\LL(G)$ is said to be {\it controllable} \cite{CY} (with respect to
$\LL(G)$ and $E_{uc}$) if $$\K\E_{uc}\cap\LL(G)\subseteq\K,$$ or equivalently,
$\K(sa)=\K(s)\wedge\LL(G)(sa)$ for any $s\in E^*$ and $a\in E_{uc}$.\label{Dfcont}
\end{Def}

We remark that a similar concept, called fuzzy controllability
condition which is the same as the above when considered for the
same underlying constituents of fuzzy automata and uncontrollable
event set, has been introduced independently by Qiu in \cite{Q05}.

It is clear that the fuzzy languages $\OO$ and $\LL(G)$ are always
controllable with respect to $\LL(G)$ and any $E_{uc}$. A good
property of controllable fuzzy languages is that they are closed
under arbitrary unions and intersections, respectively \cite{CY}.
Let $\K\subseteq\LL(G)$ be a nonempty fuzzy language. It has been
proved in \cite{CY} that there exists a (fully observable) fuzzy
supervisor $S$ for $G$ such that $\LL^S=\K$ if and only if $\K$ is
controllable.

\section{Observability}
Up to this point it has been assumed that all of the events in a fuzzy DES can be directly observed
by the fuzzy supervisor. However, in practice we usually only have local or partial observations.

To model a fuzzy DES with partial observations, we bring an additional event set $E_o\subseteq E$
of observable events that can be seen by the fuzzy supervisor, and a natural projection
$P:E\rightarrow E_o\cup\{\epsilon\}$, as usual in crisp DES. More formally, the event set $E$ is
partitioned into two disjoint subsets: $E=E_o\cup E_{uo}$, where $E_o$ and $E_{uo}$ denote the
observable event set and the unobservable event set, respectively. The natural projection $P$ is
defined in the same way as crisp DES, that is,
\begin{displaymath}
P(a)=\left\{ \begin{array}{ll}
a, & \textrm{if $a\in E_o$}\\
\epsilon, & \textrm{if $a\in E_{uo}$}.
\end{array} \right.
\end{displaymath}
The action of the projection $P$ is extended to strings by defining $P(\epsilon)=\epsilon$, and
$P(sa)=P(s)P(a)\mbox{ for any }s\in E^*\mbox{ and }a\in E$.

Due to the presence of $P$, the fuzzy supervisor cannot distinguish
between two strings $s$ and $s'$ that have the same projection,
namely, $P(s)=P(s')$. For such pairs, the fuzzy supervisor will
necessarily issue the same control action. To capture this fact, we
define a {\it partially observable fuzzy supervisor} as a map
$S_P:P[\mbox{supp}(\LL(G))]\rightarrow\F(E)$ satisfying $S(s)(a)=1$
for any $s\in P[\mbox{supp}(\LL(G))]$ and $a\in E_{uc}$. This means
that the control action may change unless an observable event
occurs. We shall assume that the control action is instantaneously
(i.e. before any unobservable event occurs) updated once an
observable event occurs. The assumption is necessary as we may wish
to update the control action for some of these unobservable events.

The behavior of $S_P/G$ when $S_P$ is controlling $G$ is defined
analogously to the case of full observation.

\vspace{0.2cm}
\begin{Def}The fuzzy language $\LL^{S_P}$ generated by $S_P/G$ is defined inductively as follows:

1) $\LL^{S_P}(\epsilon)=1;$

2) $\LL^{S_P}(sa)=\LL(G)(sa)\wedge S_P[P(s)](a)\wedge\LL^{S_P}(s)$ for any $s\in E^*$ and $a\in E$.
\end{Def}

\vspace{0.2cm} Consequently, the possibility of an event $a$
following a string $s$ under a partially observable fuzzy supervisor
only depends on the physically possible degree of $sa$ in the
controlled system, the enabled degree of $a$ after the occurrence of
$P(s)$, and  the physically possible degree of $s$ in the
closed-loop system. Clearly, $\LL^{S_P}\subseteq\LL(G)$ and
$\LL^{S_P}\in\F\LL$. For simplicity, we sometimes write $\LL$ for
$\LL(G)$.

Before introducing the concepts of observability and strong
observability from the view of event strings, we remark that based
on state estimation,  Lin and Ying have already introduced the
same terms with different essence to fuzzy DES in \cite{LY01} and
\cite{LY02}. They have used observability measure to measure the
observability of a fuzzy DES, which is a good approach to
describing the global property of the system. For our purpose, we
need a characterization of the (strong) observability of fuzzy
sublanguages.

\vspace{0.2cm}
\begin{Def}A fuzzy language $\K\subseteq\LL$ is said to be {\it observable} (respectively, {\it strongly
observable}) with respect to $\LL$, $P$, and $E_c$ if for all
$s,s'\in \mbox{supp}(\K)$ with $P(s)=P(s')$ and for any $a\in E_c$
satisfying $sa\in \mbox{supp}(\K)$, we have that
$\K(s'a)=\K(s')\wedge\LL(s'a)\wedge x$ for some (respectively,
any) $x\in\{x\in[0,1]:\K(sa)=\K(s)\wedge\LL(sa)\wedge x\}$.
\end{Def}

\vspace{0.2cm}We interpret $x$ as the enabled degrees of
controllable events when $\LL$ is restricted to $\K$. For
simplicity,  we shall omit the range $[0,1]$ of $x$ in the sequel.
Intuitively, for observable fuzzy language, we require that for
each controllable event, there is an enabled degree which can be
used after seeing the two strings that are the same under the
supervisor.

\begin{Rem}The parameter $E_c$ is included in the definition in order for the property of
observability not to overlap with the property of controllability. As we will see, observability
will only be used in conjunction with controllability, and the latter ensures that the requirements
of the definition hold for $a\in E_{uc}$.
\end{Rem}

In what follows, (strong) observability will always be with
respect to $\LL$, $P$, and $E_c$; controllability is with respect
to $\LL$ and $E_{uc}$. We simply call $\K$ (strongly) observable
or controllable if the context is clear. Note that we do not make
any specific assumptions about the relation between the
controllability and observability properties of an event. Thus, an
unobservable event could be controllable, an uncontrollable event
could be observable, and so forth.

\begin{Rem}In terms of crisp languages where membership grades are either $1$ or $0$, the above observability
as well as strong observability is the same as the observability
in  \cite{LW88}.
\end{Rem}

\begin{Rem} Clearly, $\LL$ and
$\OO$ are always observable. It is obvious by definition that the
concept of strong observability implies that of observability. But
an observable fuzzy language need not to be strongly observable.
For example, let $$E=\{a,b\}, \ \ E_c=E_o=\{b\},$$ and
$$\LL=\frac{1}{\epsilon}+\frac{0.8}{a}+\frac{0.9}{b}+\frac{0.7}{ab}.$$
One can easily check that $\LL$ is observable, but not strongly
observable.
\end{Rem}

Intuitively, a strongly observable fuzzy language $\K$ allows the associated fuzzy supervisor to
take more flexible control action than that of an observable fuzzy language, in order to exactly
achieve $\K$.  The cost of such flexibility is more constraint on $\K$ itself.

Let us provide some equivalent characterizations of observability.
\vspace{0.2cm}
\begin{Prop}For a fuzzy language $\K$, the following are equivalent:

1) \ $\K$ is observable.

2) \ For any $s\in \mbox{supp}(\K)$ and $a\in E_c$, there exists $x$
such that $\K(s'a)=\K(s')\wedge\LL(s'a)\wedge x$ for any
$s'\in\{s'\in \mbox{supp}(\K):P(s')=P(s)\}$.

3) \ For all $s,s'\in \mbox{supp}(\K)$ with $P(s)=P(s')$, and any
$a\in E_c$, there exists $x$ such that both
$\K(sa)=\K(s)\wedge\LL(sa)\wedge x$ and
$\K(s'a)=\K(s')\wedge\LL(s'a)\wedge x$.\label{Pobser}
\end{Prop}
\begin{proof}See Appendix II.
\end{proof}

\vspace{0.2cm}There is a difference between 2) and 3) above: in
2), we need for each controllable event, a common enabled degree
for all the strings that are the same under the supervisor; in 3),
we only need a common enabled degree for every two strings  that
are
 identical under the supervisor.

For strong observability, we have an equivalent characterization
too.\vspace{0.2cm}
\begin{Prop}A fuzzy language $\K$ is strongly observable if and only if for any $s,s'\in \mbox{supp}(\K)$
and $a\in E_c$ subject to $P(s)=P(s')$ and $sa,s'a\in
\mbox{supp}(\LL)$, the following conditions are satisfied:

1) \ $\K(sa)=\K(s)\wedge\LL(sa)$ if and only if $\K(s'a)=\K(s')\wedge\LL(s'a)$;

2) \ $\K(sa)=\K(s'a)$.\label{Pstrobser}
\end{Prop}
\begin{proof}See Appendix II.
\end{proof}

\vspace{0.2cm}The following results show us that (strongly)
observable fuzzy languages are closed under intersection.
\vspace{0.2cm}
\begin{Prop}\

1) \ If $\K_1$ and $\K_2$ are observable, then so is $\K_1\cap\K_2$.

2) \ If $\K_1$ and $\K_2$ are strongly observable, then so is $\K_1\cap\K_2$.\label{PInter}
\end{Prop}
\begin{proof}See Appendix II.
\end{proof}

\vspace{0.2cm}The above proposition and proof remain true for
arbitrary intersections.

\begin{Rem}Note that if $\K_1$ and $\K_2$ are strongly observable, we cannot obtain that $\K_1\cup\K_2$
is (strongly) observable in general. The following counter-example serves:

Let $E=E_c=\{a,b\}$, $E_o=\{b\}$,
$$\LL=\frac{1}{\epsilon}+\frac{0.9}{a}+\frac{0.8}{b}+\frac{0.7}{ab},$$
and $$\K_1=\frac{1}{\epsilon}+\frac{0.8}{a},\ \K_2=\frac{1}{\epsilon}+\frac{0.7}{b}.$$Then
$$\K_1\cup\K_2=\frac{1}{\epsilon}+\frac{0.8}{a}+\frac{0.7}{b}.$$It is evident that $\K_1$ and $\K_2$ are
strongly observable; however, $\K_1\cup\K_2$ is not (strongly) observable.\label{Runion}
\end{Rem}

We are now ready to present the existence result for a partially
observable fuzzy supervisor. \vspace{0.2cm}
\begin{Thm}Let $\K\subseteq\LL(G)$, where $\K$ is a nonempty fuzzy language. Then there exists a
partially observable fuzzy supervisor $S_P$ for $G$ such that
$\LL^{S_P}=\K$ if and only if $\K$ is controllable and
observable.\label{TCO}
\end{Thm}
\begin{proof}See Appendix II.
\end{proof}

\vspace{0.2cm}The proof of Theorem \ref{TCO} there is constructive
in the sense that if the controllability and observability
conditions are satisfied, it gives us a partially observable fuzzy
supervisor that will achieve the specification. For illustrating the
theorem and its proof, let us examine a simple example.

\vspace{0.2cm}
\begin{Eg}Let $E=\{a,b,c,d\}$, $E_{uc}=\{d\}$, and $E_{uo}=\{c\}$. The fuzzy languages $\LL$ and $\K$ are given by
$$\LL=\frac{1}{\epsilon}+\frac{0.9}{a}+\frac{0.8}{ab}+\frac{0.8}{ad}+\frac{0.6}{ac}+\frac{0.4}{acb}+\frac{0.6}{acd},$$
$$\K=\frac{1}{\epsilon}+\frac{0.7}{a}+\frac{0.4}{ac}+\frac{0.7}{ad}+\frac{0.4}{acd}.\qquad\qquad\qquad$$
One can check by definitions that $\K$ is controllable and
observable.  The following partially observable fuzzy supervisor
$S_P$ that follows from the definition of $S_P[P(s)]$ given in the
proof of Theorem \ref{TCO} can achieve $\K$:
$$S_P(\epsilon)=S_P[P(\epsilon)]=\frac{0.7}{a}+\frac{0}{b}+\frac{0}{c}+\frac{1}{d},\qquad\qquad\qquad$$
$$S_P(a)=S_P[P(a)]=S_P[P(ac)]=\frac{0}{a}+\frac{0}{b}+\frac{0.4}{c}+\frac{1}{d},$$
$$S_P(ab)=S_P[P(ab)]=S_P[P(acb)]=\frac{0}{a}+\frac{0}{b}+\frac{0}{c}+\frac{1}{d},$$
$$S_P(ad)=S_P[P(ad)]=S_P[P(acd)]=\frac{0}{a}+\frac{0}{b}+\frac{0}{c}+\frac{1}{d}.$$
Let us check the membership grades of $S_P(a)$ as an example:

$S_P[a](d)=1$ since $d\in E_{uc}$,

$S_P[a](a)=\vee\{\K(s'a):P(s')=a\}=0$,

$S_P[a](b)=\vee\{\K(s'b):P(s')=a\}=\K(ab)\vee\K(acb)=0$,

$S_P[a](c)=\vee\{\K(s'c):P(s')=a\}=\K(ac)=0.4$.\hfill$\blacksquare$
\end{Eg}

\vspace{0.2cm}If a fuzzy language $\K$ does not satisfy the
conditions of Theorem \ref{TCO}, then it is natural to consider
the possibility of approximating $\K$. Let us first consider the
existence of the ``least" controllable and observable fuzzy
superlanguage of $\K$. To this end, define
$$\mathscr{CO}(\K)=\{\M\in\F\LL:\K\subseteq\M\subseteq\LL\mbox{ and $\M$}\qquad\qquad$$$\qquad\qquad\qquad\qquad\qquad\qquad$
is controllable and observable$\}.$

Observe that $\LL\in\mathscr{CO}(\K)$, so the class
$\mathscr{CO}(\K)$ is not empty. Further, define
$\K^{\downarrow(CO)}=\bigcap\limits_{\M\in\mathscr{CO}(\K)}\M$.
Then from Proposition \ref{PInter} we have the following.

\vspace{0.2cm}
\begin{Prop}
The fuzzy language $\K^{\downarrow(CO)}$ is the least controllable and observable fuzzy
superlanguage of $\K$.\label{PLeastCO}
\end{Prop}
\begin{proof}Note that arbitrary intersections of controllable and observable fuzzy languages are again controllable and
observable. As a result, the proposition evidently holds.
\end{proof}

\vspace{0.2cm}Since by definition $\K^{\downarrow(CO)}\subseteq\M$
for any $\M\in\mathscr{CO}(\K)$, we call $\K^{\downarrow(CO)}$ the
{\it infimal controllable and observable fuzzy superlanguage} of
$\K$. If $\K$ is controllable and observable, then
$\K^{\downarrow(CO)}=\K$. In the ``worst" case,
$\K^{\downarrow(CO)}=\LL$.

We end this section with the Supervisory Control Problem (SCP) for fuzzy DES with partial
observation.

SCP: Given a fuzzy DES $G$ with event set $E$, uncontrollable event
set $E_{uc}\subseteq E$, observable event set $E_{o}\subseteq E$,
and two fuzzy languages $\LL_a$ and $\LL_l$, where
$\OO\neq\LL_a\subseteq\LL_l\subseteq\LL(G)$, find a partially
observable fuzzy supervisor $S_P$ such that
$\LL_a\subseteq\LL^{S_P}\subseteq\LL_l$.

Here, $\LL_a$ describes the minimal acceptable behavior and
$\LL_l$ describes the maximal legal behavior. SCP requires to find
a fuzzy supervisor such that the behavior of controlled system is
both acceptable and legal. The following result provides an
abstract solution to SCP. \vspace{0.2cm}
\begin{Coro}
There exists a partially observable fuzzy supervisor $S_P$ such that
$\LL_a\subseteq\LL^{S_P}\subseteq\LL_l$ if and only if
$\LL_a^{\downarrow(CO)}\subseteq\LL_l$.\label{Cparob}
\end{Coro}
\begin{proof}See Appendix II.
\end{proof}

\vspace{0.2cm}Although controllable fuzzy languages are closed
under arbitrary unions, observable fuzzy languages are not closed
under union in general. Consequently, it turns out that, in a
given situation, a unique maximal controllable and observable
sublanguage of $\K$ need not exist. To obtain an approximation
using sublanguages in a reasonable manner, we will introduce a new
subclass of fuzzy languages, called normal fuzzy languages, which
is deferred to Appendix I since it is not necessary for
subsequently discussing the decentralized control of fuzzy DES.

\section{Decentralized Control}
In this section, we turn our attention to the decentralized
supervisory control of a fuzzy DES that is physically distributed.
Without loss of generality and for the sake of convenience, we
consider the case that only two local partially observable fuzzy
supervisors are used to realize the decentralized supervisory
control.

The problem of decentralized control is formalized as follows. We
have two partially observable fuzzy supervisors $S_1$ and $S_2$ (for
simplicity, in the sequel we write $S$ for $S_P$), each associated
with a different projection $P_i,i=1,2$, jointly controlling the
given system $G$ with event set $E$. Associated with $G$ are the
four usual sets $E_c$, $E_{uc}$, $E_o$, and $E_{uo}$. Corresponding
to fuzzy supervisors $S_1$ and $S_2$, we have:
\begin{itemize}
    \item the sets of controllable events $E_{1c},E_{2c}\subseteq E_{c}$ satisfying $E_{1c}\cup E_{2c}=E_{c}$;
    \item the sets of observable events $E_{1o},E_{2o}\subseteq E_{o}$ satisfying $E_{1o}\cup E_{2o}=E_{o}$;
    \item the natural projection $P_i:E^*\rightarrow E^*_{io}$ corresponding to $E_{io}$ for
    $i=1,2$.
\end{itemize}
Then the local partially observable fuzzy supervisor $S_i$, $i=1,2$,
is given by $S_i:P_i[\mbox{supp}(\LL(G))]\rightarrow\F(E)$ which
satisfies $S_i(s)(a)=1$ for any $s\in P_i[\mbox{supp}(\LL(G))]$ and
$a\in E\backslash E_{ic}$.

The behavior of the controlled system under the joint supervision of $S_1$ and $S_2$ is the fuzzy
language $\LL^{S_1\wedge S_2}$ defined inductively by

1) $\LL^{S_1\wedge S_2}(\epsilon)=1$;

2) $\LL^{S_1\wedge S_2}(sa)=\LL(sa)\wedge\LL^{S_1\wedge S_2}(s)\wedge S_1[P_1(s)](a)\wedge
S_2[P_2(s)](a)$ for any $s\in E^*$ and $a\in E$.

Clearly, it follows from the definition of $\LL^{S_1\wedge S_2}$ that $\LL^{S_1\wedge
S_2}\subseteq\LL(G)$ and $\LL^{S_1\wedge S_2}\in\F\LL$.

Given a desired fuzzy language $\K\subseteq\LL(G)$, our aim is to
find the necessary and sufficient condition on $\K$ that will
ensure the existence of $S_1$ and $S_2$ such that $\LL^{S_1\wedge
S_2}=\K$. To this end, the concept of co-observability for fuzzy
languages, as a generalization of co-observability for crisp
languages \cite{RW92}, is required in place of observability
appearing in the case of centralized control.\vspace{0.2cm}

\begin{Def}A fuzzy language $\K\subseteq\LL$ is called {\it co-observable} (with respect to $\LL$,
$P_i$, and $E_{ic}$, $i=1,2$), if for all $s\in \mbox{supp}(\K)$
and $a\in E_c=E_{1c}\cup E_{2c}$, the following hold:
\begin{enumerate}
    \item if $a\in E_{1c}\cap E_{2c}$, then
$\K(sa)=\K(s)\wedge\LL(sa)\wedge[\vee_{P_1(s_1)=P_1(s)}\K(s_1a)]\wedge[\vee_{P_2(s_2)=P_2(s)}\K(s_2a)]$;
    \item if $a\in E_{1c}\backslash E_{2c}$, then
$\K(sa)=\K(s)\wedge\LL(sa)\wedge[\vee_{P_1(s_1)=P_1(s)}\K(s_1a)]$;
    \item if $a\in E_{2c}\backslash E_{1c}$, then
$\K(sa)=\K(s)\wedge\LL(sa)\wedge[\vee_{P_2(s_2)=P_2(s)}\K(s_2a)]$.
\end{enumerate}\label{Dco-ob}
\end{Def}

\vspace{0.2cm}From the above definition, it is clear that the
co-observability of fuzzy languages which takes the respective
enabled degree of controllable events under two supervisors into
account is a generation of observability.

\begin{Rem}
Like observability, co-observability is also closed under arbitrary intersections; this can be
easily verified by using the above definition. It is not difficult to check that co-observable
fuzzy languages are not closed under union in general.
\end{Rem}

The property of co-observability describes the class of fuzzy
languages that can be achieved by decentralized supervisory
control in the situation of partial observation, as shown in the
following.\vspace{0.2cm}

\begin{Thm}Let $\K\subseteq\LL(G)$ be a nonempty fuzzy language.
Then there exist two local partially observable fuzzy supervisors
$S_1$ and $S_2$ for $G$ such that $\LL^{S_1\wedge S_2}=\K$ if and
only if $\K$ is controllable and co-observable.\label{Tdc}
\end{Thm}
\begin{proof} See Appendix II.
\end{proof}

\vspace{0.2cm}The above theorem and its proof can be easily
extended for more than two local fuzzy supervisors. Analogous to
that of Theorem \ref{TCO}, the proof of Theorem \ref{Tdc} is also
constructive, and we can obtain two supervisors if the given fuzzy
sublanguage is controllable and co-observable. We present an
example to illustrate the theorem in the next section.

\section{An Illustrative Example}
In this section, we apply the previous results to an example
arising from medical diagnosis and treatment.

Suppose that there is a patient infected by a new infectious
disease. The director decides to hold a consultation first, and then
appoints one or two physicians to the patient's
physicians-in-charge. All the physicians have no complete knowledge
about the disease, but the physicians by their experience believe
that two antibiotic drugs, say penicillin and chlortetracycline, may
be useful to the disease. It is well known among medical doctors
that the drug combination of penicillin and chlortetracycline may be
have as little effect against an infection as prescribing no
antibiotic drug at all, event if the bacteria are susceptible to
each of these drugs. So the physicians decide to use the drugs
separately. Moreover, the physicians think that the  effect of
penicillin is better than that of chlortetracycline for a patient
who is in a poor condition and vice versa for a patient who is in a
fair condition, since penicillin is a bactericidal antimicrobial
drug, while chlortetracycline is a bacteriostatic one. In addition,
the drug-resistance of bacteria and some possible negative symptoms
such as fever, high white blood cell count, and increased
sedimentation rate of the blood have also been taken into account by
the physicians.

Further, the physicians consider roughly the patient's condition
to be three states: ``poor", ``fair", and ``excellent", and agree
that the present treatment must be stopped to finding other
therapies once there are some negative symptoms indicating that
the patient's condition reverts to the initial situation, i.e.,
the poor state. Such a status of diagnosis and treatment can be
logically modelled via a fuzzy DES with supervisory control, in
which the drugs and the negative symptoms are thought of as
events. Suppose that for each event, the physicians by their
experience have an estimation of the transition possibility among
 states (there are many methods for estimating membership grades;
see, for example, pages 256-260 of \cite{DP80}), and suppose that
the dose of drug which is prescribed by the physician-in-charge can
affect the transition possibility of the drug event. The drug events
are considered to be controllable and observable, while the symptom
events are uncontrollable and some of them are observable only by
those physicians who are sensitive to or concerned about the
symptoms.

\begin{figure}
\includegraphics{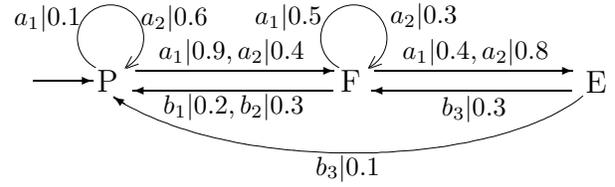}
\caption{Fuzzy automaton $G$ to model a patient's condition.}
\end{figure}

We use $a_1$ and $a_2$ to denote the drug events, namely, penicillin
and chlortetracycline. Denote by $b_i$, $i=1,2,3,$ the negative
symptoms. According to the physicians' estimation, these events and
their transition possibilities among states are depicted in Fig. 1,
where the capital letters P, F, and E represent poor, fair, and
excellent, respectively. This transition graph is a common knowledge
among the physicians after consultation. Recall that the weight of a
string is interpreted as the possibility of occurrence of the string
and is obtained by using the operation $max$ $min$. For example, the
weight of the string $a_1a_2$ is $0.8$, which is the physical
possibility to arrive at the excellent state after using penicillin
and chlortetracycline in turn.

Any therapy is required to conform to the specification which is
determined by the physicians according to their common knowledge
and the patient's situation:
$$\K=\frac{1}{\epsilon}+\frac{0.9}{a_1}+\frac{0.8}{a_1a_2}+\frac{0.2}{a_1b_1}+\frac{0.3}{a_1b_2}+\frac{0.2}{a_1a_2b_1}
+\frac{0.3}{a_1a_2b_2}$$$$+\frac{0.3}{a_1a_2b_3}+\frac{0.2}{a_1a_2b_3b_1}+\frac{0.3}{a_1a_2b_3b_2}+\frac{0.2}{a_1a_2b_3a_1}
+\frac{0.2}{a_1a_2b_3a_1b_1}$$$$+\frac{0.2}{a_1a_2b_3a_1b_2}+\frac{0.2}{a_1a_2b_3a_1b_3}+\frac{0.2}{a_1a_2b_3a_1b_3b_1}
+\frac{0.2}{a_1a_2b_3a_1b_3b_2}.$$

Now, the director appoints a physician, say $S_1$, to the patient's
physician-in-charge. We regard the physician-in-charge $S_1$ as a
supervisor and suppose that $E_{1c}=\{a_1,a_2\}$ and
$E_{1o}=\{a_1,b_1,a_2,b_3\}$. By definition, it is easy to check
that $\K$ is controllable. But $\K$ is not observable by $S_1$. In
fact, take $s=a_1, s'=a_1b_2\in \mbox{supp}(\K)$ and $a_2\in
E_{1c}$. We then see that $P_1(s)=P_1(s')=a_1$, but there is no $x$
such that both $\K(sa_2)=\K(s)\wedge\LL(sa_2)\wedge x$ and
$\K(s'a_2)=\K(s')\wedge\LL(s'a_2)\wedge x$, which contradicts 3) of
Proposition \ref{Pobser}. As a result, under the supervisory control
of $S_1$, there is no hope of achieving the specification $\K$ by
Theorem \ref{TCO}.

If the director takes possible medical errors into account and would
like to appoint one more physician, say $S_2$, to the patient's
physician-in-charge, and assume that $E_{2c}=\{a_1,a_2\}$ and
$E_{2o}=\{a_1,b_2,a_2,b_3\}$. Then the specification $\K$ can be
accomplished  through the joint control of two physicians-in-charge,
since $\K$ is now co-observable. From the proof of Theorem
\ref{Tdc}, we may take two supervisors $S_1$ and $S_2$ as follows:
$$S_1(\epsilon)=S_2(\epsilon)=\frac{0.9}{a_1}+\frac{0}{a_2}+\frac{1}{b_1}+\frac{1}{b_2}+\frac{1}{b_3},\qquad\qquad\quad$$
$$S_1(a_1)=S_2(a_1)=\frac{0}{a_1}+\frac{0.8}{a_2}+\frac{1}{b_1}+\frac{1}{b_2}+\frac{1}{b_3},\qquad\qquad$$
$$S_1(a_1a_2b_3)=S_2(a_1a_2b_3)=\frac{0.2}{a_1}+\frac{0}{a_2}+\frac{1}{b_1}+\frac{1}{b_2}+\frac{1}{b_3},\quad$$
$$S_1(\omega_1)=S_2(\omega_2)=\frac{0}{a_1}+\frac{0}{a_2}+\frac{1}{b_1}+\frac{1}{b_2}+\frac{1}{b_3}\qquad\qquad\quad$$
for any $\omega_i\in P_i[\mbox{supp}(\LL(G))]\backslash\{\epsilon,
a_1, a_1a_2b_3\}, i=1,2.$

We compute the membership grades of $S_1(a_1a_2b_3)$ and
$S_2(a_1a_2b_3)$ as an example. By the definition of $S_i[P_i(s)]$
given in the proof of Theorem \ref{Tdc}, we get that
$S_1(a_1a_2b_3)(b_j)=S_2(a_1a_2b_3)(b_j)=1$ for $j=1,2,3$, since
$b_j\in E\backslash E_{1c}$, and that
\begin{eqnarray*}
  S_1(a_1a_2b_3)(a_1)&=& \vee_{P_1(s_1)=a_1a_2b_3}\K(s_1a_1)\\
   &=&\K(a_1a_2b_3a_1)\vee
\K(a_1a_2b_3b_2a_1)\\&=&0.2\vee0=0.2,\\
S_1(a_1a_2b_3)(a_2)&=& \vee_{P_1(s_1)=a_1a_2b_3}\K(s_1a_2)\\
   &=&\K(a_1a_2b_3a_2)\vee
\K(a_1a_2b_3b_2a_2)\\&=&0\vee0=0,\\
S_2(a_1a_2b_3)(a_1)&=& \vee_{P_2(s_2)=a_1a_2b_3}\K(s_2a_1)\\
   &=&\K(a_1a_2b_3a_1)\vee
\K(a_1a_2b_3b_1a_1)\\&=&0.2\vee0=0.2,\\
S_2(a_1a_2b_3)(a_2)&=& \vee_{P_2(s_2)=a_1a_2b_3}\K(s_2a_2)\\
   &=&\K(a_1a_2b_3a_2)\vee
\K(a_1a_2b_3b_1a_2)\\&=&0\vee0=0.
\end{eqnarray*}

\section{Conclusion}
In this paper, we have established and studied the centralized and
decentralized supervisory control problem for fuzzy DES with partial
observation. Observability, normality, and co-observability of crisp
languages have been extended to fuzzy languages. We have elaborated
on the necessary and sufficient conditions on a given fuzzy language
for the existence of a partially observable fuzzy supervisor and
local fuzzy supervisors, respectively. Moreover, we have discussed
the infimal controllable and observable fuzzy superlanguage, and the
supremal controllable and normal fuzzy sublanguage. These concepts
and results are consistent with the existing theory in the framework
of Ramadge--Wonham. Moreover, we have introduced the strong
observability of fuzzy languages for comparison with the
observability.

The supervisory control formalization presented here can be
extended in many ways. Firstly, formulae and computation relating
controllability, normality, and co-observability of fuzzy
languages are yet to be established. Secondly, using
non-conjunctive fusion rules \cite{PKK97, YL02} to investigate
decentralized control of fuzzy DES is feasible. Finally,
decentralized supervisory control with local specification remains
an interesting problem.

\appendices
\section{Normality}
In this appendix, we first introduce the concept of normal fuzzy
languages which is a generalization of normality for crisp
languages in the sense of Lin and Wonham \cite{LW88D}. Then we
consider the existence of the ``largest" controllable and normal
fuzzy sublanguage, and also discuss the relation between
normality, observability, and controllability.

Recall that we have defined the natural projection
$P:E^*\rightarrow E_o^*$ whose effect on a string $s$ is to erase
the elements of $s$ that are not in $E_o$. For later need, we
extend $P$ to $\PP:\F(E^*)\rightarrow\F(E_o^*)$ via
$$\PP(\K)(\omega)=\vee_{P(\omega')=\omega}\K(\omega')$$ for any $\K\in\F(E^*)$ and $\omega\in E_o^*$.

We also define $\PP^{-1}:\F(E_o^*)\rightarrow\F(E^*)$ as follows:
$$\PP^{-1}(\M)(\omega)=\M(P(\omega))$$ for any $\M\in\F(E_o^*)$ and $\omega\in E^*$.

Clearly, the definitions of $\PP$ and $\PP^{-1}$ are well-defined;
furthermore, we have the following.\vspace{0.2cm}
\begin{Prop}Let $\PP$ and $\PP^{-1}$ be defined as above.

1) \ If $\K$ is a fuzzy language over $E$, then $\PP(\K)$ is a fuzzy
language over $E_o$.

2) \ If $\M$ is a fuzzy language over $E_o$, then $\PP^{-1}(\M)$
is a fuzzy language over $E$.\label{Pnorm}
\end{Prop}
\begin{proof} See Appendix II.
\end{proof}

\vspace{0.2cm}Recall that in \cite{LW88D} (cf. the definition of
recognizable languages in  \cite{CDFV88}) a crisp language
$K\subseteq L$ is called {\it normal} (with respect to $L$ and
$P$) if $\overline{K}=P^{-1}[P(\overline{K})]\cap L$. For
normality of fuzzy languages, we have to incorporate membership
grades into the normality of their supports as crisp languages.
Keeping the previous notation, we can now introduce the normality
of fuzzy languages.\vspace{0.2cm}
\begin{Def}
A fuzzy language $\K\subseteq\LL$ is said to be {\it normal} (with
respect to $\LL$ and $P$) if
$$\K=\PP^{-1}[\PP(\K)]\cap\LL.$$
\end{Def}

\vspace{0.2cm}Intuitively, normality for a fuzzy language means
that $\K$ can be exactly recovered from its projection $\PP(\K)$
and $\LL$ itself. Observe that both $\OO$ and $\LL$ are normal
with respect to $\LL$ and $P$, so the above property is not
vacuous.

The notion of fuzzy language normality reduces to normality of
crisp languages if membership grades in fuzzy languages are only
allowed to be either $0$ or $1$. The following relationship
between them provides a necessary condition for a fuzzy language
to be normal.\vspace{0.2cm}

\begin{Prop}If a fuzzy language $\K\subseteq\LL$ is normal with respect to $\LL$ and $P$, then
the crisp language $\mbox{supp}(\K)$ is normal with respect to
$\mbox{supp}(\LL)$ and $P$ in the sense of crisp language
normality. \label{Pnorm2}
\end{Prop}
\begin{proof}See Appendix II.
\end{proof}

\vspace{0.2cm}A desired property of normal fuzzy languages is that
they are closed under arbitrary unions, as shown
below.\vspace{0.2cm}

\begin{Prop}If for each $i\in I$, $\K_i\subseteq\LL$ is normal with respect to $\LL$ and $P$, then so is
$\bigcup_{i\in I}\K_i$.\label{Punion}
\end{Prop}
\begin{proof}See Appendix II.
\end{proof}

\vspace{0.2cm}The following theorem shows us that normality is
stronger than observability.\vspace{0.2cm}
\begin{Thm}
If $\K\subseteq\LL$ is normal with respect to $\LL$ and $P$, then
$\K$ is observable with respect to $\LL$, $P$, and $E_c$ for any
$E_c\subseteq E$.\label{Tnorm}
\end{Thm}
\begin{proof}See Appendix II.
\end{proof}
\vspace{0.2cm}
\begin{Rem}The converse statement of the above theorem is not true in general. For instance, $\K_1$ in
Remark \ref{Runion} is observable; however, it is not normal.
\end{Rem}

We are now in the position to present an approximation of $\K$
using controllable and observable sublanguages, as promised. For
this, let us introduce a new class of fuzzy sublanguages of $\K$
as follows: $$\mathscr{CN}(\K)=\{\M\subseteq\K:\M\mbox{ is
controllable and normal}\}.$$Observe that
$\OO\in\mathscr{CN}(\K)$, so the class is not empty.

Define
$\K^{\uparrow(CN)}=\bigcup\limits_{\M\in\mathscr{CN}(\K)}\M$.
Next, using Proposition \ref{Punion} we can establish the
existence of the supremal controllable and normal fuzzy
sublanguage of $\K$.\vspace{0.2cm}

\begin{Prop}The fuzzy language $\K^{\uparrow(CN)}$ is the largest controllable and normal fuzzy sublanguage of $\K$.
\end{Prop}
\begin{proof}Note that arbitrary unions of controllable and normal fuzzy languages are again controllable and
normal. Consequently, the proposition holds obviously.
\end{proof}

\vspace{0.2cm}It should be pointed out that $\K^{\uparrow(CN)}$
may not in general a maximal controllable and observable fuzzy
sublanguage of $\K$. In other words, there may be controllable and
observable fuzzy sublanguages that are strictly larger than the
supremal controllable and normal fuzzy sublanguage.

As shown earlier, an observable fuzzy language is not normal in
general; however, under certain conditions, observability does
imply normality.\vspace{0.2cm}

\begin{Thm}Suppose that $E_c\subseteq E_o$. If $\K\subseteq\LL$ is controllable (with respect to
$\LL$ and $E_{uc}$) and observable (with respect to $\LL$, $P$,
and $E_c$), then $\K$ is normal (with respect to $\LL$ and $P$).
\label{Tnorm2}
\end{Thm}
\begin{proof}See Appendix II.
\end{proof}

\vspace{0.2cm}The importance of this theorem lies in the fact that
when the assumption $E_c\subseteq E_o$ is fulfilled, that is, when
all the controllable events are observable, or equivalently when
all the unobservable events are uncontrollable, the theorem
implies that the supremal controllable and observable fuzzy
sublanguage does exist, which is given by $\K^{\uparrow(CN)}$.

\section{Proofs}
For the need of proofs, we make an observation first.

\begin{Lem}Let $\K$ be an observable fuzzy language, $s_1,s_2\in \mbox{supp}(\K)$ with $P(s_1)=P(s_2)$, and $a\in
E_c$.

1) \ If $\K(s_ia)<\K(s_i)\wedge\LL(s_ia)$ for $i=1,2$, then
$\K(s_1a)=\K(s_2a)$.

2) \ If $\K(s_1a)<\K(s_1)\wedge\LL(s_1a)$ and
$\K(s_2a)=\K(s_2)\wedge\LL(s_2a)$, then
$\K(s_2a)\leq\K(s_1a)$.\label{Lobser}
\end{Lem}
\begin{proof}

1): If $s_1a\not\in \mbox{supp}(\K)$ and $s_2a\not\in
\mbox{supp}(\K)$, then $\K(s_1a)=\K(s_2a)=0$. Otherwise, there is
$s_ia$, say $s_2a$, in $\mbox{supp}(\K)$. By the definition of
observability, there exists $x$ such that
$\K(s_1a)=\K(s_1)\wedge\LL(s_1a)\wedge x$ and
$\K(s_2a)=\K(s_2)\wedge\LL(s_2a)\wedge x$. From the assumption
that $\K(s_ia)<\K(s_i)\wedge\LL(s_ia)$ for $i=1,2$, we see that
$\K(s_1a)=x=\K(s_2a)$, as desired.

2): If $\K(s_2a)=0$, then it is obvious that
$\K(s_2a)\leq\K(s_1a)$. In the case of $\K(s_2a)>0$, there exists
$x$ such that $\K(s_1a)=\K(s_1)\wedge\LL(s_1a)\wedge x$ and
$\K(s_2a)=\K(s_2)\wedge\LL(s_2a)\wedge x$ by the definition of
observability. Hence, it follows from the conditions of 2) that
$\K(s_1a)=x$ and $\K(s_2a)\leq x$. Consequently,
$\K(s_2a)\leq\K(s_1a)$.
\end{proof}

\vspace{0.2cm} {\it Proof of Proposition \ref{Pobser}:}

1)$\Rightarrow$2): For any $s\in \mbox{supp}(\K)$ and $a\in E_c$,
set $[s]=\{s'\in \mbox{supp}(\K):P(s')=P(s)\}$ and
$x_{[s]a}=\vee_{s'\in[s]}\K(s'a)$. We claim that $x=x_{[s]a}$
satisfies $\K(s'a)=\K(s')\wedge\LL(s'a)\wedge x$ for any $s'\in[s]$.
In fact, if $\K(s'a)=\K(s')\wedge\LL(s'a)$, then the claim evidently
holds since $\K(s'a)\leq x_{[s]a}$. If
$\K(s'a)<\K(s')\wedge\LL(s'a)$, then by Lemma \ref{Lobser} we have
that $x_{[s]a}=\vee_{s'\in[s]}\K(s'a)=\K(s'a)$, and the claim holds
too.

Observe that 2)$\Rightarrow$3) and 3)$\Rightarrow$1) are obvious,
so the proposition is true. \hfill$\blacksquare$

\vspace{0.2cm}{\it Proof of Proposition \ref{Pstrobser}:} We first
prove the necessity. Let $s,s'\in \mbox{supp}(\K)$ and $a\in E_c$
such that $P(s)=P(s')$ and $sa,s'a\in \mbox{supp}(\LL)$.

For condition 1), we only prove the `only if' part; the `if' part
is symmetric. Suppose that $\K(sa)=\K(s)\wedge\LL(sa)$. From the
assumptions that $s\in \mbox{supp}(\K)$ and $sa\in
\mbox{supp}(\LL)$, we see that $\K(sa)>0$. Thus by the definition
of strong observability, we know that for any $x$, if
$\K(sa)=\K(s)\wedge\LL(sa)\wedge x$, then
$\K(s'a)=\K(s')\wedge\LL(s'a)\wedge x$. In particular, taking
$x=1$, we get that
$\K(s'a)=\K(s')\wedge\LL(s'a)\wedge1=\K(s')\wedge\LL(s'a)$.

To verify condition 2), let us first show the fact that $\K(sa)=0$
if and only if $\K(s'a)=0$. We prove the `if' part; the `only if'
part is similar. Suppose that $\K(s'a)=0$. By contradiction,
assume that $\K(sa)>0$. Then by the definition of strong
observability, for any $x\in\{x:\K(sa)=\K(s)\wedge\LL(sa)\wedge
x\}$, we must have that $\K(s'a)=\K(s')\wedge\LL(s'a)\wedge x$.
However, taking $x=\K(sa)$, we see that
$\K(s'a)=\K(s')\wedge\LL(s'a)\wedge \K(sa)>0$ since $s'\in
\mbox{supp}(\K)$ and $s'a\in \mbox{supp}(\LL)$ by the assumptions.
It is a contradiction. Next, we consider the case that both
$\K(sa)>0$ and $\K(s'a)>0$. By the foregoing property P2) we see
that $\K(sa)\leq\K(s)$ and $\K(s'a)\leq\K(s')$. Note that
$\K\subseteq\LL$ which means that $\K(\omega)\leq\LL(\omega)$ for
any $\omega$. Consequently, $\K(sa)\leq\K(s)\wedge\LL(sa)$ and
$\K(s'a)\leq\K(s')\wedge\LL(s'a)$. By 1), there are only two
cases: The first is that both $\K(sa)=\K(s)\wedge\LL(sa)$ and
$\K(s'a)=\K(s')\wedge\LL(s'a)$; the second is that both
$\K(sa)<\K(s)\wedge\LL(sa)$ and $\K(s'a)<\K(s')\wedge\LL(s'a)$.
For the first case, by taking $x=\K(sa)$ in the definition of
strong observability, we have that
$\K(s'a)=\K(s')\wedge\LL(s'a)\wedge\K(sa)$, which yields that
$\K(s'a)\leq\K(sa)$. By interchanging $s$ and $s'$, and taking
$x=\K(s'a)$, we get that $\K(sa)\leq\K(s'a)$. Hence
$\K(sa)=\K(s'a)$ in the first case. For the second case, noting
that a strongly observable fuzzy language is observable, we obtain
that $\K(sa)=\K(s'a)$ from Lemma \ref{Lobser}, thus finishing the
proof of the necessity.

Now, let us show the sufficiency. Suppose that $s,s'\in
\mbox{supp}(\K)$ with $P(s)=P(s')$, $a\in E_c$ satisfying $sa\in
\mbox{supp}(\K)$, and $x_0\in\{x:\K(sa)=\K(s)\wedge\LL(sa)\wedge
x\}$. It needs to verify that $\K(s'a)=\K(s')\wedge\LL(s'a)\wedge
x_0$.

If $\LL(s'a)=0$, then it forces that $\K(s'a)=0$, and thus
$\K(s'a)=\K(s')\wedge\LL(s'a)\wedge x_0$.

In the case of $\LL(s'a)>0$, conditions 1) and 2) can be
applicable. Further, if $\K(sa)=\K(s)\wedge\LL(sa)$, then we get
that
 $\{x\in[0,1]:\K(sa)=\K(s)\wedge\LL(sa)\wedge x\}=[\K(sa),1]$ and $\K(s'a)=\K(s')\wedge\LL(s'a)$ by condition
 1). Therefore, $x_0\geq\K(s'a)$ by condition 2), and thus $\K(s'a)=\K(s')\wedge\LL(s'a)\wedge x_0$
 holds. If $\K(sa)<\K(s)\wedge\LL(sa)$, then $\{x:\K(sa)=\K(s)\wedge\LL(sa)\wedge x\}=\{\K(sa)\}$ and
 $\K(s'a)<\K(s')\wedge\LL(s'a)$ by condition 1). Hence, $x_0=\K(s'a)$ by condition 2) and
 $\K(s'a)=\K(s')\wedge\LL(s'a)\wedge x_0$ holds. This completes the proof of the
 proposition. \hfill $\blacksquare$

\vspace{0.2cm}
 {\it Proof of Proposition \ref{PInter}:}

1) \ Suppose that $s,s'\in \mbox{supp}(\K_1\cap\K_2)$ and $a\in
E_c$ such that $P(s)=P(s')$ and $sa\in \mbox{supp}(\K_1\cap\K_2)$.
Since $\K_i$, $i=1,2$ is observable, there exists $x_i$ such that
$\K_i(sa)=\K_i(s)\wedge\LL(sa)\wedge x_i$ and
$\K_i(s'a)=\K_i(s')\wedge\LL(s'a)\wedge x_i$. Setting $x=x_1\wedge
x_2$, we can easily check that
$(\K_1\cap\K_2)(sa)=(\K_1\cap\K_2)(s)\wedge\LL(sa)\wedge x$ and
$(\K_1\cap\K_2)(s'a)=(\K_1\cap\K_2)(s')\wedge\LL(s'a)\wedge x$. It
follows from definition that $\K_1\cap\K_2$ is observable.

2) \ By contradiction, suppose that $\K_1\cap\K_2$ is not strongly
observable. Then by definition there exist $s,s'\in
\mbox{supp}(\K_1\cap\K_2)$ with $P(s)=P(s')$, $a\in E_c$, and $x$
such that $sa\in \mbox{supp}(\K_1\cap\K_2)$ and
$$\K_1(sa)\wedge\K_2(sa)=\K_1(s)\wedge\K_2(s)\wedge\LL(sa)\wedge
x,\eqno(1)$$ but
$$\K_1(s'a)\wedge\K_2(s'a)\neq\K_1(s')\wedge\K_2(s')\wedge\LL(s'a)\wedge
x.\eqno(2)$$ The last inequality implies that $s'a\in
\mbox{supp}(\LL)$; otherwise, both sides of inequality (2) are
zero. As $sa\in \mbox{supp}(\K_1\cap\K_2)$, we get that $sa\in
\mbox{supp}(\LL)$. Thus Proposition \ref{Pstrobser} is applicable
to $\K_1$ and $\K_2$. As a result, we have the following: for
$i=1,2$,
$$\K_i(sa)=\K_i(s)\wedge\LL(sa)\Leftrightarrow\K_i(s'a)=\K_i(s')\wedge\LL(s'a),\eqno(3)$$
and
$$\K_i(sa)=\K_i(s'a).\eqno(4)$$Four cases need to be discussed:

Case 1: $\K_i(s'a)=\K_i(s')\wedge\LL(s'a)$, $i=1,2$. Using (3), we
see that $\K_i(sa)=\K_i(s)\wedge\LL(sa)$. Applying them to (1)
yields that $\K_1(sa)\wedge\K_2(sa)=\K_1(sa)\wedge\K_2(sa)\wedge
x$. However, we find from (2) and (4) that
$\K_1(sa)\wedge\K_2(sa)\neq\K_1(sa)\wedge\K_2(sa)\wedge x$. This
is a contradiction.

Case 2: $\K_i(s'a)<\K_i(s')\wedge\LL(s'a)$, $i=1,2$. Using (3)
again, we have that $\K_i(sa)<\K_i(s)\wedge\LL(sa)$. Hence
$\K_1(sa)\wedge\K_2(sa)<\K_1(s)\wedge\K_2(s)\wedge\LL(sa)$, which
means that $\K_1(sa)\wedge\K_2(sa)=x$ by (1). Using (4) and
$\K_i(s'a)<\K_i(s')\wedge\LL(s'a)$, we have that
$x=\K_1(sa)\wedge\K_2(sa)=\K_1(s'a)\wedge\K_2(s'a)<\K_1(s')\wedge\K_2(s')\wedge\LL(s'a)$.
This forces that
$\K_1(s'a)\wedge\K_2(s'a)=\K_1(s')\wedge\K_2(s')\wedge\LL(s'a)\wedge
x$, which contradicts with (2).

Case 3: $\K_1(s'a)=\K_1(s')\wedge\LL(s'a)$ and
$\K_2(s'a)<\K_2(s')\wedge\LL(s'a)$. By (3), we also have that
$\K_1(sa)=\K_1(s)\wedge\LL(sa)$ and
$\K_2(sa)<\K_2(s)\wedge\LL(sa)$. In the subcase of
$\K_1(sa)\leq\K_2(sa)$, we also get that $\K_1(s'a)\leq\K_2(s'a)$
by (4). Thus
$\K_1(sa)\wedge\K_2(sa)=\K_1(s)\wedge\K_2(s)\wedge\LL(sa)$ and
$\K_1(s'a)\wedge\K_2(s'a)=\K_1(s')\wedge\K_2(s')\wedge\LL(s'a)$.
From the former and (1), we see that
$\K_1(sa)\wedge\K_2(sa)=\K_1(sa)\wedge\K_2(sa)\wedge x$; from the
latter and (2), we see that
$\K_1(s'a)\wedge\K_2('sa)\neq\K_1(s'a)\wedge\K_2(s'a)\wedge x$.
This is absurd by (4). In the other subcase, i.e.,
$\K_1(sa)>\K_2(sa)$, we see that $\K_1(s'a)>\K_2(s'a)$ from (4).
Further, we have that
$\K_1(sa)\wedge\K_2(sa)<\K_1(s)\wedge\K_2(s)\wedge\LL(sa)$ and
$\K_1(s'a)\wedge\K_2(s'a)<\K_1(s')\wedge\K_2(s')\wedge\LL(s'a)$.
The former and (1) force that $x=\K_1(sa)\wedge\K_2(sa)$.
Therefore $x=\K_1(s'a)\wedge\K_2(s'a)$, and thus
$\K_1(s')\wedge\K_2(s')\wedge\LL(s'a)\wedge
x=\K_1(s'a)\wedge\K_2(s'a)$, which contradicts with (2).

Case 4: $\K_1(s'a)<\K_1(s')\wedge\LL(s'a)$ and
$\K_2(s'a)=\K_2(s')\wedge\LL(s'a)$. This case is symmetric to Case
3, so we omit its proof. \hfill$\blacksquare$

\vspace{0.2cm} {\it Proof of Theorem \ref{TCO}:} We prove the
sufficiency first. Suppose that $\K$ is both controllable and
observable. For any $P(s)\in P[\mbox{supp}(\LL(G))]$, let us
define
\begin{displaymath}
S_P[P(s)](a)=\left\{ \begin{array}{ll}
1, & \textrm{if $a\in E_{uc}$}\\
\vee\{\K(s'a):P(s')=P(s)\}, & \textrm{if $a\in E_{c}$}.
\end{array} \right.
\end{displaymath}Clearly, with the above definition $S_P$ is a partially observable fuzzy supervisor.

It remains to verify that $\LL^{S_P}=\K$. We prove this by using
induction on the length of strings. The base case is for strings
of length $0$. By the definition of $\LL^{S_P}$, we see that
$\LL^{S_P}(\epsilon)=1=\K(\epsilon)$. So the base case holds. The
induction hypothesis is that $\LL^{S_P}(s)=\K(s)$ for all strings
$s$ with length $n$. We now prove the same for strings $sa$, where
$a\in E$. In the case of $a\in E_{uc}$, we obtain by definition
that
\begin{eqnarray*}
\LL^{S_P}(sa)
 &=&\LL(G)(sa)\wedge
S_P[P(s)](a)\wedge\LL^{S_P}(s) \\
 &=&\LL(G)(sa)\wedge\LL^{S_P}(s)\mbox{ (by definition of $S_P$)}
\\&=&\LL(G)(sa)\wedge\K(s)\mbox{ (using induction hypothesis)}
\\&=&\K(sa),\mbox{ (by controllability of $\K$)}
 \end{eqnarray*}i.e., $\LL^{S_P}(sa)=\K(sa)$. Now, let us consider the other
case $a\in E_{c}$. In this case,
\begin{eqnarray*}
\LL^{S_P}(sa)&=&\LL(G)(sa)\wedge S_P[P(s)](a)\wedge\LL^{S_P}(s) \\
&=&\LL(G)(sa)\wedge S_P[P(s)](a)\wedge\K(s)
\\&=&\LL(G)(sa)\wedge[\vee\{\K(s'a):P(s')=P(s)\}]\\& &\wedge\K(s).\hspace{5cm} (5)
\end{eqnarray*}
Thus, it is clear that $\LL^{S_P}(sa)\geq\K(sa)$. By
contradiction, suppose that $\LL^{S_P}(sa)>\K(sa)$. Then from (5)
we get that $\LL(G)(sa)\wedge\K(s)>\K(sa)$ and
$\vee\{\K(s'a):P(s')=P(s)\}>\K(sa)$. So there exists $s'$ with
$P(s')=P(s)$ such that $\K(s'a)>\K(sa)$.

If $\K(sa)>0$, then it is necessary that $s,s'\in
\mbox{supp}(\K)$. Furthermore, by the observability of $\K$, there
exists $x$ such that both $\K(sa)=\K(s)\wedge\LL(G)(sa)\wedge x$
and $\K(s'a)=\K(s')\wedge\LL(G)(s'a)\wedge x$. The first equality,
together with the previous argument that
$\LL(G)(sa)\wedge\K(s)>\K(sa)$, forces that $x=\K(sa)$; the second
equality implies that $x\geq\K(s'a)$. Consequently,
$\K(sa)\geq\K(s'a)$, which contradicts with $\K(s'a)>\K(sa)$.

In the case of $\K(sa)=0$, if $\vee\{\K(s'a):P(s')=P(s)\}=0$, then
we see from (5) that $\LL^{S_P}(sa)=0=\K(sa)$; otherwise, there
exists $s'$ such that $\K(s'a)>0$ and $P(s')=P(s)$. Clearly,
$s'\in \mbox{supp}(\K)$. Note also that $s\in \mbox{supp}(\K)$
since $\LL(G)(sa)\wedge\K(s)>\K(sa)$. Again, by the observability
of $\K$, there exists $x$ such that both
$\K(s'a)=\K(s')\wedge\LL(G)(s'a)\wedge x$ and
$\K(sa)=\K(s)\wedge\LL(G)(sa)\wedge x$. The second equality,
together with the fact that $\LL(G)(sa)\wedge\K(s)>\K(sa)$, forces
that $x=\K(sa)=0$. Therefore
$\K(s'a)=\K(s')\wedge\LL(G)(s'a)\wedge x=0$, which contradicts
with $\K(s'a)>0$. This completes the proof of the induction step.

Next, to see the necessity, suppose that there exists a partially
observable fuzzy supervisor $S_P$ for $G$ such that $\LL^{S_P}=\K$.
For controllability, by definition it suffices to show that
$\LL^{S_P}(sa)=\LL^{S_P}(s)\wedge\LL(G)(sa)$ for any $s\in E^*$ and
$a\in E_{uc}$. In fact, by definition we have that
\begin{eqnarray*}\LL^{S_P}(sa)&=&\LL^{S_P}(s)\wedge
S_P[P(s)](a)\wedge\LL(G)(sa)\\&=&\LL^{S_P}(s)\wedge\LL(G)(sa),\end{eqnarray*}
since $S_P[P(s)](a)=1$ for each $a\in E_{uc}$. To prove the
observability of $\K$, take any $s,s'\in \mbox{supp}(\K)$ with
$P(s)=P(s')$ and $a\in E_c$ such that $\K(sa)>0$. Selecting
$x=S_P[P(s)](a)$, we have that
\begin{eqnarray*}
\K(sa)&=&\LL^{S_P}(sa)\\
 &=&\LL^{S_P}(s)\wedge\LL(G)(sa)\wedge S_P[P(s)](a) \\
 &=&\K(s)\wedge\LL(G)(sa)\wedge x
\end{eqnarray*}and
\begin{eqnarray*}
\K(s'a)&=&\LL^{S_P}(s'a)\\
&=&\LL^{S_P}(s')\wedge\LL(G)(s'a)\wedge S_P[P(s')](a) \\
&=&\K(s')\wedge\LL(G)(s'a)\wedge S_P[P(s)](a) \\
&=&\K(s')\wedge\LL(G)(s'a)\wedge x.
\end{eqnarray*}
Hence, $\K$ is observable. The proof of the theorem is completed.
\hfill$\blacksquare$

\vspace{0.2cm}{\it Proof of Corollary \ref{Cparob}:} Suppose that
there is a partially observable fuzzy supervisor $S_P$ such that
$\LL_a\subseteq\LL^{S_P}\subseteq\LL_l$. Then by Theorem \ref{TCO}
we know that $\LL^{S_P}\supseteq\LL_a$ is both controllable and
observable. Note that $\LL_a^{\downarrow(CO)}$ is the infimal
controllable and observable fuzzy superlanguage of $\LL_a$ by
Proposition \ref{PLeastCO}, so
$\LL_a^{\downarrow(CO)}\subseteq\LL^{S_P}\subseteq\LL_l$.

Conversely, assume that $\LL_a^{\downarrow(CO)}\subseteq\LL_l$.
Clearly, the controllable and observable fuzzy language
$\LL_a^{\downarrow(CO)}$ is not the empty fuzzy language. Therefore
by Theorem \ref{TCO} there is a partially observable fuzzy
supervisor $S_P$ such that $\LL^{S_P}=\LL_a^{\downarrow(CO)}$, and
thus $\LL_a\subseteq\LL^{S_P}\subseteq\LL_l$. \hfill $\blacksquare$

\vspace{0.2cm}{\it Proof of Theorem \ref{Tdc}:} We first prove the
necessity. Suppose that there exist two local partially observable
fuzzy supervisors $S_1$ and $S_2$ for $G$ such that $\LL^{S_1\wedge
S_2}=\K$. To see the controllability of $\K$, let $s\in E^*$ and
$a\in E_{uc}$. Then we obtain that
\begin{eqnarray*}
\K(sa)&=&\LL^{S_1\wedge S_2}(sa)\quad \mbox{(by assumption)}\\
&=&\LL(sa)\wedge\LL^{S_1\wedge S_2}(s)\wedge S_1[P_1(s)](a)\\
& &\wedge S_2[P_2(s)](a)\quad \mbox{(by definition)}\\
&=&\LL(sa)\wedge\LL^{S_1\wedge S_2}(s).\quad \mbox{(since $a\in
E_{uc}$)}
\end{eqnarray*}
Hence, $\K$ is controllable by the definition of controllability.

Now we consider the co-observability. Let $s\in \mbox{supp}(\K)$
and $a\in E_c$. The assumption $\K=\LL^{S_1\wedge S_2}$ gives rise
to
$$\K(sa)=\LL(sa)\wedge\LL^{S_1\wedge S_2}(s)\wedge S_1[P_1(s)](a)\wedge S_2[P_2(s)](a).\eqno(6)$$
From this equality, we always have that
$\K(sa)\leq\LL(sa)\wedge\LL^{S_1\wedge S_2}(s)$.  Note that if
$\K(sa)=\LL(sa)\wedge\LL^{S_1\wedge S_2}(s)$, then all the
conditions 1), 2), and 3) in the definition of co-observability
are satisfied since $\vee_{P_1(s_1)=P_1(s)}\K(s_1a)\geq\K(sa)$ and
$\vee_{P_2(s_2)=P_2(s)}\K(s_2a)\geq\K(sa)$. It remains to discuss
the case of $\K(sa)<\LL(sa)\wedge\LL^{S_1\wedge S_2}(s)$. In this
case, we see from (6) that
$$\K(sa)=S_1[P_1(s)](a)\wedge S_2[P_2(s)](a).\eqno(7)$$Let us consider all possible subcases:

Subcase 1: $a\in E_{1c}\cap E_{2c}$. According to Definition
\ref{Dco-ob}, we need to show that
$\K(sa)=\K(s)\wedge\LL(sa)\wedge[\vee_{P_1(s_1)=P_1(s)}\K(s_1a)]\wedge[\vee_{P_2(s_2)=P_2(s)}\K(s_2a)]$.
Observe that
$\K(sa)\leq\K(s)\wedge\LL(sa)\wedge[\vee_{P_1(s_1)=P_1(s)}\K(s_1a)]\wedge[\vee_{P_2(s_2)=P_2(s)}\K(s_2a)]$.
Assume that
$\K(sa)<\K(s)\wedge\LL(sa)\wedge[\vee_{P_1(s_1)=P_1(s)}\K(s_1a)]\wedge[\vee_{P_2(s_2)=P_2(s)}\K(s_2a)]$.
Then we obtain that $\K(sa)<\vee_{P_i(s_i)=P_i(s)}\K(s_ia)$ for
$i=1,2$. So there exists $s'_i$ with $P_i(s'_i)=P_i(s)$, $i=1,2$,
such that $\K(s'_ia)>\K(sa)$.

On the other hand, from the assumption $\K=\LL^{S_1\wedge S_2}$
and $P_i(s'_i)=P_i(s)$, we have that
\begin{eqnarray*}
\K(s'_ia)&=&\LL^{S_1\wedge S_2}(s'_ia)\\
&=&\LL(s'_ia)\wedge\LL^{S_1\wedge S_2}(s'_i)\wedge S_1[P_1(s'_i)](a)\\& &\wedge S_2[P_2(s'_i)](a)\\
&\leq&S_1[P_1(s'_i)](a)\wedge S_2[P_2(s'_i)](a)\\
&\leq&S_i[P_i(s)](a),
\end{eqnarray*}that is, $\K(s'_ia)\leq S_i[P_i(s)](a)$ for $i=1,2$.
Further, if $S_1[P_1(s)](a)\leq S_2[P_2(s)](a)$, then (7) forces
that $\K(sa)=S_1[P_1(s)](a)$. Consequently, $\K(s'_1a)\leq\K(sa)$,
which contradicts with the foregoing argument $\K(s'_ia)>\K(sa)$. If
$S_1[P_1(s)](a)> S_2[P_2(s)](a)$, then (7) yields that
$\K(sa)=S_2[P_2(s)](a)$. This, together with $\K(s'_ia)\leq
S_i[P_i(s)](a)$, leads to $\K(s'_2a)\leq\K(sa)$, which is again a
contradiction. So the previous assumption that
$\K(sa)<\K(s)\wedge\LL(sa)\wedge[\vee_{P_1(s_1)=P_1(s)}\K(s_1a)]\wedge[\vee_{P_2(s_2)=P_2(s)}\K(s_2a)]$
does not work, and we have thus proved Subcase 1.

Subcase 2: $a\in E_{1c}\backslash E_{2c}$. Now we need to verify
that
$\K(sa)=\K(s)\wedge\LL(sa)\wedge[\vee_{P_1(s_1)=P_1(s)}\K(s_1a)]$.
Since $a\not\in E_{2c}$, by definition we have that
$S_2[P_2(s)](a)=1$. Accordingly, we see from (7) that
$\K(sa)=S_1[P_1(s)](a)$. Clearly,
$\K(sa)\leq\K(s)\wedge\LL(sa)\wedge[\vee_{P_1(s_1)=P_1(s)}\K(s_1a)]$.
Similar to the proof of Subcase 1, we assume that
$\K(sa)<\K(s)\wedge\LL(sa)\wedge[\vee_{P_1(s_1)=P_1(s)}\K(s_1a)]$,
which means that $\K(sa)<\vee_{P_1(s_1)=P_1(s)}\K(s_1a)$.
Therefore there exists $s'_1$ with $P_1(s'_1)=P_1(s)$ such that
$\K(s'_1a)>\K(sa)$. Notice that
\begin{eqnarray*}
\K(s'_1a)&=&\LL^{S_1\wedge S_2}(s'_1a)\\
&=&\LL(s'_1a)\wedge\LL^{S_1\wedge S_2}(s'_1)\wedge S_1[P_1(s'_1)](a)\\& &\wedge S_2[P_2(s'_1)](a)\\
&\leq&S_1[P_1(s'_1)](a)\\
&=&S_1[P_1(s)](a),
\end{eqnarray*}i.e., $\K(s'_1a)\leq S_1[P_1(s)](a)$. This, together with $\K(sa)=S_1[P_1(s)](a)$,
gives rise to $\K(s'_1a)\leq\K(sa)$, which contradicts with
$\K(s'_1a)>\K(sa)$. Thus Subcase 2 is proved.

Subcase 3: $a\in E_{2c}\backslash E_{1c}$. It is similar to
Subcase 2, so we can omit its proof. So far we have completed the
proof of the necessity.

Next, we show the sufficiency  by constructing two partially
observable fuzzy supervisors
$S_i:P_i[\mbox{supp}(\LL(G))]\rightarrow\F(E)$. Suppose that $\K$ is
controllable and co-observable. For any $s\in \mbox{supp}(\LL(G))$,
define $S_i[P_i(s)]$ according to
\begin{displaymath}
S_i[P_i(s)](a)=\left\{ \begin{array}{ll}
1, & \textrm{if $a\in E\backslash E_{ic}$}\\
\vee_{P_i(s_i)=P_i(s)}\K(s_ia), & \textrm{if $a\in E_{ic}$.}
\end{array} \right.
\end{displaymath} Obviously, $S_i$, $i=1,2$ is a fuzzy supervisor.

We claim that $\LL^{S_1\wedge S_2}=\K$, which will be proved by
induction on the length of string $s$.

Basis step: $s=\epsilon$. Clearly, $\LL^{S_1\wedge
S_2}(\epsilon)=1=\K(\epsilon)$ by the definition of
$\LL^{S_1\wedge S_2}$ and the condition $\K\neq\OO$. So the basis
step is valid.

Induction step: Assume that for all strings $s$ with $|s|\leq n$,
we have that $\LL^{S_1\wedge S_2}(s)=\K(s)$. We now prove the same
for strings of the form $sa$, where $a\in E$. Four possible cases
need to be considered.

Case 1: $a\in E_{uc}$. By the definition of $\LL^{S_1\wedge S_2}$,
we have that
\begin{eqnarray*}
\LL^{S_1\wedge S_2}(sa)&=&\LL(sa)\wedge\LL^{S_1\wedge S_2}(s)\wedge S_1[P_1(s)](a)\\
& &\wedge S_2[P_2(s)](a)\\
&=&\LL(sa)\wedge\LL^{S_1\wedge S_2}(s)\quad \mbox{(since $a\in E_{uc}$)}\\
&=&\LL(sa)\wedge\K(s)\ \mbox{(by induction hypothesis)}\\
&=&\K(sa),\quad \mbox{(by controllability of $\K$)}
\end{eqnarray*}i.e., $\LL^{S_1\wedge S_2}(sa)=\K(sa)$, as desired.

Case 2: $a\in E_{1c}\cap E_{2c}$. In this case, the
co-observability of $\K$ can be applicable, and we obtain that
\begin{eqnarray*}
\LL^{S_1\wedge S_2}(sa)&=&\LL(sa)\wedge\LL^{S_1\wedge S_2}(s)\wedge S_1[P_1(s)](a)\\
& &\wedge S_2[P_2(s)](a)\\
&=&\LL(sa)\wedge\K(s)\wedge S_1[P_1(s)](a)\\
& &\wedge S_2[P_2(s)](a)\ \mbox{(by induction hypothesis)}\\
&=&\LL(sa)\wedge\K(s)\wedge[\vee_{P_1(s_1)=P_1(s)}\K(s_1a)]\\
& &\wedge[\vee_{P_2(s_2)=P_2(s)}\K(s_2a)]\ \mbox{(by definition of
$S_i$)}
\\&=&\K(sa),\quad
\mbox{(by co-observability of $\K$)}
\end{eqnarray*}i.e., $\LL^{S_1\wedge S_2}(sa)=\K(sa)$.

Case 3: $a\in E_{1c}\backslash E_{2c}$. In this case, the
co-observability of $\K$ can also be applicable, and we have that
\begin{eqnarray*}
\LL^{S_1\wedge S_2}(sa)&=&\LL(sa)\wedge\LL^{S_1\wedge S_2}(s)\wedge S_1[P_1(s)](a)\\
& &\wedge S_2[P_2(s)](a)\\
&=&\LL(sa)\wedge\K(s)\wedge S_1[P_1(s)](a)\ \mbox{(since $a\not\in E_{2c}$)}\\
&=&\LL(sa)\wedge\K(s)\wedge[\vee_{P_1(s_1)=P_1(s)}\K(s_1a)]\\
& &\qquad\qquad\qquad\qquad \mbox{(by definition of $S_1$)}
\\&=&\K(sa),\quad
\mbox{(by co-observability of $\K$)}
\end{eqnarray*}i.e., $\LL^{S_1\wedge S_2}(sa)=\K(sa)$.

Case 4: $a\in E_{2c}\backslash E_{1c}$. It is similar to Case 3,
and the proof is omitted. This finishes the proof of the theorem.
\hfill $\blacksquare$

\vspace{0.2cm}{\it Proof of Proposition \ref{Pnorm}:}

1): If $\K$ is the empty fuzzy language, then it is evident that
$\PP(\K)$ is the empty fuzzy language. For $\K\neq\OO$, we need to
check the properties P1) and P2) in the definition of fuzzy
languages. Obviously,
$\PP(\K)(\epsilon)=\vee_{P(\omega')=\epsilon}\K(\omega')=1$ since
$\K(\epsilon)=1$. Hence, P1) holds. To see P2), it is sufficient
to show that $\PP(\K)(sa)\leq\PP(\K)(s)$ for any $s\in E_o^*$ and
$a\in E_o$. By contradiction, assume that $\PP(\K)(sa)>\PP(\K)(s)$
for some $s\in E_o^*$ and $a\in E_o$. Note that
$\PP(\K)(sa)=\vee_{P(\omega')=sa}\K(\omega')$ and
$\PP(\K)(s)=\vee_{P(\mu')=s}\K(\mu')$. Thereby, there exists
$\omega'_0\in E^*$ with $P(\omega'_0)=sa$ such that
$\K(\omega'_0)>\K(\mu')$ for any $\mu'\in E^*$ satisfying
$P(\mu')=s$. Since $P(\omega'_0)=sa$, we can write $\omega'_0$ as
$\omega'_1\omega'_2$ which satisfies $P(\omega'_1)=s$ and
$P(\omega'_2)=a$. Thus, we get that
$\K(\omega'_1\omega'_2)>\K(\omega'_1)$, a contradiction. We
therefore conclude that P2) holds.

2): If $\M=\OO$, then it is clear that $\PP^{-1}(\M)=\OO$. For
$\M\neq\OO$, we check the properties P1) and P2). By definition,
we see that
$\PP^{-1}(\M)(\epsilon)=\M(P(\epsilon))=\M(\epsilon)=1$. So P1)
holds. For P2), it suffices to verify that
$\PP^{-1}(\M)(sa)\leq\PP^{-1}(\M)(s)$, namely $\M(P(sa))\leq
\M(P(s))$, for any $s\in E^*$ and $a\in E$. If $a\in E_{uo}$, then
$P(sa)=P(s)$, and thus $\M(P(sa))=\M(P(s))$. If $a\in E_{o}$, then
$P(sa)=P(s)a$, and thus $\M(P(sa))=\M(P(s)a)\leq \M(P(s))$. This
completes the proof. \hfill $\blacksquare$

\vspace{0.2cm}{\it Proof of Proposition \ref{Pnorm2}:} By
definition and the fact that $\mbox{supp}(\K)$ is closed, we need
to show that $\mbox{supp}(\K)=P^{-1}[P(\mbox{supp}(\K))]\cap
\mbox{supp}(\LL)$. It is obvious that $\mbox{supp}(\K)\subseteq
P^{-1}[P(\mbox{supp}(\K))]\cap \mbox{supp}(\LL)$, so we need only
to prove that $P^{-1}[P(\mbox{supp}(\K))]\cap
\mbox{supp}(\LL)\subseteq \mbox{supp}(\K)$. Suppose that $s\in
P^{-1}[P(\mbox{supp}(\K))]\cap \mbox{supp}(\LL)$. Then $\LL(s)>0$
and $P(s)\in P(\mbox{supp}(\K))$. So there exists $s'\in
\mbox{supp}(\K)$ such that $P(s')=P(s)$. Thus, from the assumption
that $\K$ is normal, we have that
\begin{eqnarray*}
\K(s)&=&[\PP^{-1}[\PP(\K)]\cap\LL](s)\\
 &=&\PP^{-1}[\PP(\K)](s)\wedge\LL(s)\\
 &=&[\vee_{P(\omega')=P(s)}\K(\omega')]\wedge\LL(s)
 \\&\geq&\K(s')\wedge\LL(s)
 \\&>&0,
 \end{eqnarray*}namely, $s\in \mbox{supp}(\K)$. Thus, $P^{-1}[P(\mbox{supp}(\K))]\cap
\mbox{supp}(\LL)\subseteq \mbox{supp}(\K)$, finishing the proof.
\hfill $\blacksquare$

\vspace{0.2cm}{\it Proof of Proposition \ref{Punion}:} For
simplicity, we only prove the case that $I$ has two elements.
There is no difficulty to generalize the proof to infinite index
set. Suppose that $\K_1$ and $\K_2$ are normal fuzzy languages,
namely, $\K_i=\PP^{-1}[\PP(\K_i)]\cap\LL$ for $i=1,2.$ To prove
the normality of $\K_1\cup\K_2$, it suffices to show that
$(\K_1\cup\K_2)(s)=\PP^{-1}[\PP(\K_1\cup\K_2)](s)\wedge\LL(s)$ for
any $s\in E^*$. By the normality of $\K_i$, we obtain that
\begin{eqnarray*}
\K_i(s)&=&[\PP^{-1}[\PP(\K_i)]\cap\LL](s)\\
 &=&\PP^{-1}[\PP(\K_i)](s)\wedge\LL(s)\\
 &=&[\vee_{P(\omega')=P(s)}\K_i(\omega')]\wedge\LL(s)
\end{eqnarray*}for $i=1,2$. Therefore, for any $s\in E^*$ we have the following:
\begin{eqnarray*}
(\K_1\cup\K_2)(s)&=&\K_1(s)\vee\K_2(s)\\
&=&[(\vee_{P(\omega')=P(s)}\K_1(\omega'))\wedge\LL(s)]\vee\\&
&\qquad\qquad
[(\vee_{P(\omega')=P(s)}\K_2(\omega'))\wedge\LL(s)]\\
&=&[(\vee_{P(\omega')=P(s)}\K_1(\omega'))\vee\\&
&\qquad(\vee_{P(\omega')=P(s)}\K_2(\omega'))]\wedge\LL(s)
\\&=&[\vee_{P(\omega')=P(s)}(\K_1(\omega')\vee\K_2(\omega'))]\wedge\LL(s)
\\&=&[\vee_{P(\omega')=P(s)}(\K_1\cup\K_2)(\omega')]\wedge\LL(s)
\\&=&\PP^{-1}[\PP(\K_1\cup\K_2)](s)\wedge\LL(s),
\end{eqnarray*}i.e., $(\K_1\cup\K_2)(s)=\PP^{-1}[\PP(\K_1\cup\K_2)](s)\wedge\LL(s)$, thus finishing
the proof of the proposition. \hfill $\blacksquare$

\vspace{0.2cm}{\it Proof of Theorem \ref{Tnorm}:} We use the
statement 3) of Proposition \ref{Pobser} to show the observability
of $\K$. Let $s,s'\in \mbox{supp}(\K)$ with $P(s)=P(s')$, and
$a\in E_c$. It suffices to prove the claim that
$\K(sa)=\K(s)\wedge\LL(sa)\wedge x$ and
$\K(s'a)=\K(s')\wedge\LL(s'a)\wedge x$ have a common solution.  By
the definition of normality, we have the following:
\begin{eqnarray*}
\K(sa)&=&[\PP^{-1}[\PP(\K)]\cap\LL](sa)\\
 &=&\PP^{-1}[\PP(\K)](sa)\wedge\LL(sa)\\
 &=&[\vee_{P(\omega')=P(sa)}\K(\omega')]\wedge\LL(sa)
 \\&=:&\lambda\wedge\LL(sa)\mbox{ (write $\lambda$ for $\vee_{P(\omega')=P(sa)}\K(\omega')$)}
 \end{eqnarray*}and
 \begin{eqnarray*}
\K(s'a)&=&[\PP^{-1}[\PP(\K)]\cap\LL](s'a)\\
 &=&\PP^{-1}[\PP(\K)](s'a)\wedge\LL(s'a)\\
 &=&[\vee_{P(\omega')=P(s'a)}\K(\omega')]\wedge\LL(s'a)
 \\&=&[\vee_{P(\omega')=P(sa)}\K(\omega')]\wedge\LL(s'a)
 \\&=&\lambda\wedge\LL(s'a).
 \end{eqnarray*}
To give a common solution $x$, let us consider all possible cases.

Case 1: $\K(sa)=\K(s)\wedge\LL(sa)$ and
$\K(s'a)=\K(s')\wedge\LL(s'a)$. In this case, the claim evidently
holds by taking $x=1$.

Case 2: $\K(sa)=\K(s)\wedge\LL(sa)$ and
$\K(s'a)<\K(s')\wedge\LL(s'a)$. In this case, we first verify that
$\K(sa)\leq\K(s'a)$. From the previous arguments, we find that
$\lambda\wedge\LL(s'a)=\K(s'a)<\K(s')\wedge\LL(s'a)\leq\LL(s'a)$.
We thus get that $\lambda<\LL(s'a)$, and furthermore,
$\lambda=\K(s'a)$. Note
 that $\K(sa)=\lambda\wedge\LL(sa)\leq\lambda$. Therefore, $\K(sa)\leq\K(s'a)$. Taking $x=\K(s'a)$,
 we see that the claim holds.

Similarly, the claim holds for $\K(sa)<\K(s)\wedge\LL(sa)$ and
$\K(s'a)=\K(s')\wedge\LL(s'a)$.

Case 3: $\K(sa)<\K(s)\wedge\LL(sa)$ and
$\K(s'a)<\K(s')\wedge\LL(s'a)$. In this case, we have that
$\lambda\wedge\LL(sa)=\K(sa)<\K(s)\wedge\LL(sa)\leq\LL(sa)$.
Consequently, $\lambda<\LL(sa)$, and thus $\lambda=\K(sa)$. In the
same way, we can get that $\lambda=\K(s'a)$. Thus the claim holds
by taking $x=\lambda$. The proof of the theorem is finished.
\hfill $\blacksquare$

\vspace{0.2cm}{\it Proof of Theorem \ref{Tnorm2}:} If $\K$ is the
empty fuzzy language, then the result is trivial. Now, let us
consider the case of $\K\neq\OO$. Observe that
$\K\subseteq\PP^{-1}[\PP(\K)]\cap\LL$, so we need only to prove
that $\PP^{-1}[\PP(\K)]\cap\LL\subseteq\K$. It is sufficient to
verify that $[\PP^{-1}[\PP(\K)]\cap\LL](s)\leq\K(s)$ for any $s\in
E^*$. By contradiction, assume that there exists $t\in E^*$ such
that $[\PP^{-1}[\PP(\K)]\cap\LL](t)>\K(t)$. It is clear that
$t\neq\epsilon$, since $\K\neq\OO$. Let $sa$ be the shortest such
$t$, that is, $[\PP^{-1}[\PP(\K)]\cap\LL](sa)>\K(sa)$ but
$[\PP^{-1}[\PP(\K)]\cap\LL](\omega)=\K(\omega)$ for any $\omega\in
E^*$ with $|\omega|\leq|s|$. Two cases need to be discussed.

Case 1: $a\in E_{uc}$. By the controllability of $\K$, we get that
$\K(sa)=\K(s)\wedge\LL(sa)$. From the previous assumption, we
obtain that
\begin{eqnarray*}\K(s)&=&\PP^{-1}[\PP(\K)](s)\wedge\LL(s)
\\&=&[\vee_{P(\mu')=P(s)}\K(\mu')]\wedge\LL(s).
\end{eqnarray*}Hence, we get by $\LL(s)\geq\LL(sa)$ that
\begin{eqnarray*}\K(sa)&=&\K(s)\wedge\LL(sa)\\&=&[\vee_{P(\mu')=P(s)}\K(\mu')]\wedge\LL(s)\wedge\LL(sa)
\\&=&[\vee_{P(\mu')=P(s)}\K(\mu')]\wedge\LL(sa).
\end{eqnarray*} In the subcase of $a\in E_{uo}$, we thus have that
\begin{eqnarray*}
[\PP^{-1}[\PP(\K)]\cap\LL](sa)&=&\PP^{-1}[\PP(\K)](sa)\wedge\LL(sa)\\
 &=&[\vee_{P(\omega')=P(sa)}\K(\omega')]\wedge\LL(sa)
 \\&=&[\vee_{P(\omega')=P(s)}\K(\omega')]\wedge\LL(sa)
 \\&=&\K(sa).
\end{eqnarray*}This contradicts with the assumption that $[\PP^{-1}[\PP(\K)]\cap\LL](sa)>\K(sa)$.
In the other subcase $a\in E_o$, we get that
$[\PP^{-1}[\PP(\K)]\cap\LL](sa)=[\vee_{P(\omega')=P(s)a}\K(\omega')]\wedge\LL(sa)$.
Therefore, the assumption $[\PP^{-1}[\PP(\K)]\cap\LL](sa)>\K(sa)$
implies that
$[\vee_{P(\omega')=P(s)a}\K(\omega')]\wedge\LL(sa)>[\vee_{P(\mu')=P(s)}\K(\mu')]\wedge\LL(sa),$
which forces that
$\vee_{P(\omega')=P(s)a}\K(\omega')>\vee_{P(\mu')=P(s)}\K(\mu')$.
Whence, there exists $\omega'_0\in E^*$ with $P(\omega'_0)=P(s)a$
such that $\K(\omega'_0)>\K(\mu')$ for all $\mu'\in E^*$
satisfying $P(\mu')=P(s)$. As a result, we can write $\omega'_0$
as $\omega'_{01}\omega'_{02}$, where $P(\omega'_{01})=P(s)$ and
$P(\omega'_{02})=a$, and furthermore, we see that
$\K(\omega'_{01}\omega'_{02})>\K(\omega'_{01})$. This is absurd.

Case 2: $a\in E_c$. From the condition $E_c\subseteq E_o$, we see
that $a\in E_o$. Hence the assumption
$[\PP^{-1}[\PP(\K)]\cap\LL](sa)>\K(sa)$ implies that
$[\vee_{P(\omega')=P(s)a}\K(\omega')]\wedge\LL(sa)>\K(sa)$. So
there exists $\omega'\in E^*$ with $P(\omega')=P(s)a$ such that
$\K(\omega')\wedge\LL(sa)>\K(sa)$. Since $P(\omega')=P(s)a$, we
can set $\omega'=s'a\omega''$ such that $P(s')=P(s)$ and
$P(\omega'')=\epsilon$. We thus see that $\K(s'a)\geq\K(\omega')$,
and moreover, $\K(s'a)\wedge\LL(sa)>\K(sa)$. The latter forces
that both $\K(s'a)>\K(sa)$ and $\LL(sa)>\K(sa)$.

Because we need Lemma \ref{Lobser} in the later development, let
us pause to check its conditions. We have known that $P(s')=P(s)$
and $a\in E_c$, so it remains only to show that $s,s'\in
\mbox{supp}(\K)$. In fact, by the previous arguments we have that
$\K(s)=[\PP^{-1}[\PP(\K)]\cap\LL](s)\geq[\PP^{-1}[\PP(\K)]\cap\LL](sa)>\K(sa)\geq0$,
i.e., $\K(s)>0$, and $\K(s')\geq\K(s'a)>\K(sa)\geq0$, i.e.,
$\K(s')>0$.

From $\K(s'a)>\K(sa)$ and Lemma \ref{Lobser}, we obtain that
$$\K(sa)=\K(s)\wedge\LL(sa)\eqno(8)$$ and
$$\K(s'a)\leq\K(s')\wedge\LL(s'a).\eqno(9)$$  From (8) and the
proven fact $\LL(sa)>\K(sa)$, we find that
$\K(s)=\K(sa)<\LL(sa)\leq\LL(s)$, i.e., $\K(s)<\LL(s)$. Note that
$\K(s)=[\vee_{P(\mu')=P(s)}\K(\mu')]\wedge\LL(s)$, so we get that
$\K(s)=\vee_{P(\mu')=P(s)}\K(\mu')$. On the other hand, we see
from $\K\subseteq\PP^{-1}[\PP(\K)]\cap\LL$ that
\begin{eqnarray*}
\K(s')&\leq&[\PP^{-1}[\PP(\K)]\wedge\LL](s')
\\&=&[\vee_{P(\omega')=P(s')}\K(\omega')]\wedge\LL(s')\\
&=&[\vee_{P(\omega')=P(s)}\K(\omega')]\wedge\LL(s')
\\&\leq&\vee_{P(\omega')=P(s)}\K(\omega')\\
&=&\K(s),
\end{eqnarray*}i.e., $\K(s')\leq\K(s)$. Accordingly,
$\K(s'a)\leq\K(s')\leq\K(s)=\K(sa)$, that is, $\K(s'a)\leq\K(sa)$,
which contradicts with the proven fact $\K(s'a)>\K(sa)$. Thus the
proof of the theorem is completed. \hfill$\blacksquare$

\end{document}